\documentclass[prd,twocolumn,floatfix,amsmath,nofootinbib,amssymb,floatfix]{revtex4}
\usepackage{graphicx,color,dcolumn,booktabs,bm}
\usepackage{longtable,lscape}
\usepackage{pdfpages}
\usepackage{txfonts}
\usepackage{overpic}
\usepackage{amssymb}
\usepackage{makecell}
\usepackage{indentfirst}
\usepackage{feynmf}   
\usepackage{slashed}  
\usepackage{cases}
\usepackage{color}
\usepackage{multirow}
\usepackage{threeparttable}
\usepackage{epstopdf}
\usepackage{enumerate}
\usepackage{subfigure}
\usepackage{diagbox}
\usepackage{graphicx,color,dcolumn,booktabs,bm}
\usepackage[colorlinks,
            citecolor=blue,
            anchorcolor=red,
            menucolor=red,
            linkcolor=red,
            filecolor=red,
            runcolor=red,
            urlcolor=blue,
            frenchlinks=red]{hyperref}

\begin{document}
\title{Systematic analysis of doubly charmed baryons $\Xi_{cc}$ and $\Omega_{cc}$}
\author{Guo-Liang Yu$^{1}$}
\email{yuguoliang2011@163.com}
\author{Yan Meng$^{1}$}
\author{Zhen-Yu Li$^{2}$}
\email{zhenyvli@163.com}
\author{Zhi-Gang Wang$^{1}$}
\email{zgwang@aliyun.com}
\author{Lu Jie$^{1}$}

\affiliation{$^1$ Department of Mathematics and Physics, North China
Electric Power University, Baoding 071003, People's Republic of
China\\$^2$ School of Physics and Electronic Science, Guizhou Education University, Guiyang 550018, People's Republic of
China}
\date{\today }

\begin{abstract}
In this work, we perform a systematic study of the mass spectra, the root mean square(r.m.s.) radii and the radial density distributions of the doubly charmed baryons $\Xi_{cc}$ and $\Omega_{cc}$. The calculations are carried out in the frame work of relativized quark model, where the baryon is regarded as a three-body system of quarks. Our results show that the excited energy of doubly charmed baryon with $\rho$-mode is lower than those of the $\lambda$-mode and $\lambda$-$\rho$ mixing mode, which indicates that the lowest state is dominated by the $\rho$-mode. According to this conclusion, we systematically investigate the mass spectra, the r.m.s. radii of the ground and excited states($1S\sim4S$, $1P\sim4P$, $1D\sim4D$, $1F\sim4F$ and $1G\sim4G$) with $\rho$-mode. Using the wave functions obtained from quark model, we also study the radial density distributions. Finally, with the predicated mass spectra, the Regge trajectories of $\Xi_{cc}$ and $\Omega_{cc}$ in the ($J$,$M^{2}$) plane are constructed, and the slopes, intercepts are determined by linear fitting. It is found that model predicted masses fit nicely to the constructed Regge trajectories.
\end{abstract}


\maketitle

\section{Introduction}

The investigation of doubly heavy baryons is of great interest to experimental and theoretical physicist, as it provides a good opportunity for us to understand the strong interactions and basic QCD theory.
Up to now, many single heavy baryons have been well discovered by Belle, BABAR, CLEO and LHCb collaborations\cite{PDG} and the mass spectra of single heavy baryons have become more and more abundance. However, searching for doubly heavy baryons in experiments ended with no progress for a long time.
The first observation of a doubly heavy baryon $\Xi_{cc}^{+}$(3519) was reported by the SELEX collaboration in 2002 in the decay mode $\Xi_{cc}^{+}\rightarrow \Lambda_{c}^{+}K^{-}\pi^{+}$\cite{35191}. Although SELEX collaboration confirmed this state in another decay mode $pD^{+}K^{-}$\cite{35192}, the FOCUS, BaBar, Belle and LHCb collaborations reported no evidence of the production of this doubly charmed baryon\cite{35193,35194,35195,35196}. The breakthrough came in 2017 with the discovery of a doubly charmed baryon $\Xi_{cc}^{++}$ by the LHCb collaboration\cite{352117}. This state was observed in the decay mode $\Xi_{cc}^{++}\rightarrow \Lambda_{c}^{+}K^{-}\pi^{+}\pi^{+}$ with a measured mass $3621.40\pm0.72\pm0.14\pm0.27$ MeV and later was confirmed in another decay mode $\Xi_{cc}^{++}\rightarrow \Lambda_{c}^{+}\pi^{+}$\cite{352118,352120}.

In theory, the mass spectra of the doubly heavy baryons have been predicted with various methods, such as the relativistic or nonrelativistic quark model\cite{QM0,QM1,QM2,QM4,QM5,QM7,QM8,QM9,QM10,QM11,QM12,QM13,QM14,QM15,QM17,QM18,QM19,QM21,QM22,QM23,QM25,QM27,QM34,QM35,QM36,QM37,QM38,QM39}, QCD sum rules\cite{Sum1,Sum2,Sum3,Sum4,Sum5,Sum6,Sum7,Sum8,Sum9}, bag models\cite{Bag1,Bag2,Bag3,Bag4}, the Bethe-Salpeter equation\cite{BSE1,BSE2,BSE3,BSE4}, effective field theories\cite{EFT1,EFT2,EFT3,EFT4}, Lattice QCD\cite{lattice1,lattice2,lattice3,lattice4,lattice5,lattice6,lattice7} and the others\cite{Other1,Other2,Other3,Other4,Other5}. To our knowledge, only Refs.\cite{QM2,QM25} focused on the mass spectra of the doubly heavy baryons from the ground states to the high excited states systematically in the quark-diquark picture. Under this picture, the initial three-body problem is reduced to two-step two-body calculations. However, the popular quark-diquark picture of a baryon is not universal and its results needs further confirmation by different methods. Thus, it is necessary for us give a systematic analysis of the properties of ground and excited states of doubly heavy baryons.

The relativized quark model, developed first by Godfrey, Capstick and Isgur\cite{GI1,GI2}, has been widely used to investigate the properties of the mesons, baryons, and evenly the tetraquark states\cite{LV1,LV2,LV3}. In this model, the relativistic effects are involved, which may be essential for doubly heavy baryon involving a light quark. Since the baryon is a three-body system, its theory is much more complicated compared to the two-body meson system, especially in the calculations of the matrix elements of the Hamiltonian in quark model. In our previous work, we employed a method of infinitesimally-shifted Gaussian(ISG) basis function in the relativized quark model\cite{GLY1,ZYL1}, where the calculation of the matrix element is simplified and the baryon is treated as a real three-body system.

In the present work, we use the method in Refs.\cite{GLY1,ZYL1} to study the mass spectra and r.m.s. radii of the excited doubly charmed baryons up to rather high orbital and radial excitations. With the predicted mass spectra, we construct the Regge trajectories in the ($J$,$M^{2}$) planes and determine their Regge slopes and intercepts. Using the wave functions obtained from the relativized quark model, we also study the radial density distributions of the doubly charmed baryons. The paper is organized as follows. After the introduction, we briefly describe the phenomenological methods adopted in this work in Sec.II. In Sec.III we present our numerical results and discussions about $\Xi_{cc}$ and $\Omega_{cc}$. In Sec.IV the baryon Regge trajectories in the ($J$, $M^{2}$) plane are constructed. And Sec V is reserved for our conclusions.
\begin{figure*}[htbp]
\centering
\includegraphics[height=4.5cm,width=16cm]{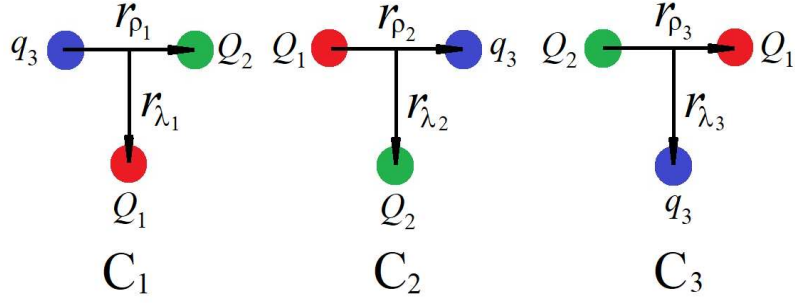}
\caption{ Jacobi coordinates for the three-body system.}
\label{figure0}
\end{figure*}
\section{ Phenomenological methods adopted in this work}
\subsection{ Wave function of doubly charmed baryon}

The doubly charmed baryon is a three-body system which contains two charmed quarks and one light quark($u$, $d$ or $s$ quark) inside. The inter-quark interaction in this three-body system is commonly described by the three sets of Jacobi coordinates in Fig. \ref{figure0}. Each set of Jacobi coordinate is called a channel($c$) and is defined as,
\begin{eqnarray}
& \boldsymbol{r}_{\lambda_{i}}=\textbf{r}_{i}-\frac{m_{j}\textbf{r}_{i}+m_{k}\textbf{r}_{k}}{m_{j}+m_{k}} & \\
& \boldsymbol{r}_{\rho_{i}}=\textbf{r}_{j}-\textbf{r}_{k}&
\end{eqnarray}
where $i$, $j$, $k$=1, 2, 3 (or replace their positions in turn). $\mathbf{r}_{i}$ and $m_{i}$ denote the position vector and
the mass of the $i$th quark, respectively.

In the heavy quark limit, one light quark within the doubly charmed baryon is decoupled from two heavy quarks. It can be seen from Fig. \ref{figure0} that channel $3$ properly reflects the characteristic of the heavy quark symmetry.
Thus, the calculations in this work are performed based on channel 3. Using the transformation
of Jacobi coordinates, we can calculate all the matrix elements in channel 3.
Under this picture, the degree of freedom between two heavy quarks is commonly called the $\rho$-mode, while the degree between the center of mass of two heavy quarks and the light quark is called the $\lambda$-mode.
It was indicated by Refs.\cite{QM22,GLY1,ZYL1} that the lowest state of a single heavy baryon is dominated by the $\lambda$-mode. We will see in the following analysis that the doubly charmed baryons are dominated by $\rho$-mode.

In this work, we employ Gaussian basis functions\cite{Gaussian1} to construct the orbital part of the wave function for a three-body system, which can be written as,
\begin{eqnarray}
\phi_{nlm_{l}}(\boldsymbol{r})=N_{nl}r^{l}e^{-\nu_{n}r^{2}}Y_{lm_{l}}(\hat{\boldsymbol{r}}), \quad (n=1-n_{max})
\end{eqnarray}
with
\begin{eqnarray}
N_{nl}=\sqrt{\frac{2^{l+2}(2\nu_{n})^{l+3/2}}{\sqrt{\pi}(2l+1)!!}}
\end{eqnarray}
\begin{eqnarray}
\nu_{n}=\frac{1}{r_{n}^{2}}, \quad r_{n}=r_{a}\Big[\frac{r_{amax}}{r_{a}}\Big]^{\frac{n-1}{n_{max}-1}}
\end{eqnarray}
In Eq.(5), $n_{max}$ is the maximum number of the Gaussian basis functions, $r_{a}$ and $r_{amax}$ are the Gaussian range parameters. In different studies, people employed different values for $r_{a}$ and $r_{amax}$ such as $r_{a}$=0.05$\sim$0.3 fm and $r_{amax}$=10$\sim$15 fm in Ref.\cite{Gaussian2}, and $r_{a}$=0.1 fm and $r_{amax}$=20 fm in Ref.\cite{Gaussian3}. The three parameters $n_{max}$, $r_{a}$ and $r_{amax}$ are actually related to each other. As illustrated in Figs. \ref{figure0-1} and \ref{figure0-2} in Sec. 3.1, the results show well stability when the parameters are taken as $r_{a}$=0.18 fm, $r_{amax}$=15 fm with $n_{max}=9\sim12$. The orbital wave function is constructed from the wave functions of the two Jacobi coordinates $\rho$ and $\lambda$, and takes the form,
\begin{eqnarray}
\Phi_{l_{\rho},l_{\lambda},L}=\big[\phi_{n_{\rho}l_{\rho}m_{l_{\rho}}}(\boldsymbol{r}_{\rho})\phi_{n_{\lambda}l_{\lambda}m_{l_{\lambda}}}(\boldsymbol{r}_{\lambda})\big]_{L}
\end{eqnarray}
The spatial part of the wave function includes the spin wave function and orbital part, which can be written as,
\begin{align}
\psi_{JM}=\big[\big[[\chi_{1/2}(Q)\chi_{1/2}(Q)]_{s}\Phi_{l_{\rho},l_{\lambda},L}\big]_{j}\chi_{1/2}(q)\big]_{JM}
\end{align}
where $\chi_{1/2}$ is the spin wave function of quark and $\textbf{\emph{s}}$ is the total spin of two heavy quarks. In the heavy quark limit, the coupling scheme of the spin
and angular momenta is as, $\textbf{\emph{L}}=\textbf{\emph{l}}_{\rho}+\textbf{\emph{l}}_{\lambda}$, $\textbf{\emph{j}}=\textbf{\emph{s}}+\textbf{\emph{L}}$ and $\textbf{\emph{J}}=\textbf{\emph{j}}+\frac{1}{2}$.
Finally, the full wave function for a definite state of a baryon can be expressed as,
\begin{align}
\Psi_{full}^{JM}=\sum_{n_{_{\rho}},n_{_{\lambda}}}C_{n_{_{\rho}},n_{_{\lambda}}}\Psi_{JM}(\boldsymbol{r}_{\rho},\boldsymbol{r}_{\lambda})
 \quad (n_{\rho},n_{\lambda}=1,\cdots,n_{max})
\end{align}
where $\Psi_{JM}(\boldsymbol{r}_{\rho},\boldsymbol{r}_{\lambda})$ is the direct product of color wave function, flavor wave function and the spatial wave function
\begin{eqnarray}
\Psi_{JM}(\boldsymbol{r}_{\rho},\boldsymbol{r}_{\lambda})=\phi_{\mathrm{color}}\otimes \phi_{\mathrm{flavor}}\otimes\psi_{JM}
\end{eqnarray}
The state of a doubly charmed baryon can be characterized by a given quantum numbers $l_{\rho}$, $l_{\lambda}$, $L$, $s$, $j$ and $J^{P}$.

The flavor wave function and color function of a doubly charmed baryon is symmetric and antisymmetric, respectively. We know that the total wave function must be antisymmetric, thus the spatial part should always be symmetric. For a double quark system in a baryon, its spin wave function is antisymmetric singlet($s=0$) or symmetric triplet($s=1$). Correspondingly, the orbital part must also be antisymmetric or symmetric to couple a symmetric spatial wave function. Thus, the total spin $s$ of two charmed quarks($cc$) and orbital quantum number $l_{\rho}$ should satisfy the condition $(-1)^{s+l_{\rho}} = -1$.

\subsection{ The relativistic quark model and ISG method}

In the relativistic quark model, baryons are formed by three valence(constituent) quarks. They are confined by a confining potential and interact with each
other by residual two-body interactions. In the framework of relativistic quark model, the Hamiltonian for a three-body system is of the form\cite{GI1,GI2},
\begin{align}
\widehat{H}=\sum_{i=1}^{3}(p_{i}^{2}+m_{i}^{2})^{1/2}+\sum_{i<j}\widetilde{H}_{ij}^{\mathrm{conf}}+\sum_{i<j}\widetilde{H}_{ij}^{\mathrm{hyp}}+\sum_{i<j}\widetilde{H}_{ij}^{\mathrm{so}}
\end{align}
where the first term is the relativistic kinetic energy term, $\widetilde{H}^{\mathrm{conf}}_{ij}$ is the spin-independent potential including a linear confining potential $\widetilde{S}(r_{ij})$ and the one-gluon
exchange potential $G^{\prime}(r_{ij})$,
\begin{eqnarray}
\widetilde{H}^{\mathrm{conf}}_{ij}=\widetilde{S}(r_{ij})+G^{\prime}(r_{ij})
\end{eqnarray}
The linear confining potential $\widetilde{S}(r_{ij})$ can be written as,
\begin{eqnarray}
\widetilde{S}(r_{ij})=-\frac{3}{4}\textbf{\emph{F}}_{i}\cdot\textbf{\emph{F}}_{j}\Big\{b r_{ij}\Big[\frac{e^{-\sigma_{ij}^{2}r_{ij}^{2}}}{\sqrt{\pi}\sigma_{ij} r_{ij}} \notag \\
+\big(1+\frac{1}{2\sigma_{ij}^{2}r_{ij}^{2}}\big)\frac{2}{\sqrt{\pi}} \int^{\sigma_{ij} r_{ij}}_{0}e^{-x^{2}}dx\Big]+c\Big\}
\end{eqnarray}
with
\begin{align}
&\sigma_{ij}=\sqrt{s^{2}\Big[\frac{2m_{i}m_{j}}{m_{i}+m_{j}}\Big]^{2}+\sigma_{0}^{2}\Big[\frac{1}{2}\big(\frac{4m_{i}m_{j}}{(m_{i}+m_{j})^{2}}\big)^{4}+\frac{1}{2}\Big]} &&
\end{align}
In Eq.(12), $\textbf{\emph{F}}_{i}\cdot\textbf{\emph{F}}_{j}$ stands for the color matrix and $F_{n}$ reads,
\begin{equation}
F_{n}=\left\{
      \begin{array}{l}
       \frac{\lambda_{n}}{2} \quad \mathrm{for} \, \mathrm{quarks}, \\
        -\frac{\lambda_{n}^{*}}{2} \quad    \mathrm{for} \, \mathrm{antiquarks} \\
      \end{array}
      \right.
\end{equation}
with $n=1,2\cdots8$. The one-gluon exchange potential $G^{\prime}(r_{ij})$ can be expressed in terms of one-gluon-exchange propagator $\widetilde{G}(r_{ij})$,
\begin{eqnarray}
G^{\prime}(r_{ij})=\Big(1+\frac{p^{2}_{ij}}{E_{i}E_{j}}\Big)^{\frac{1}{2}}\widetilde{G}(r_{ij})\Big(1+\frac{p^{2}_{ij}}{E_{i}E_{j}}\Big)^{\frac{1}{2}}
\end{eqnarray}
with
\begin{eqnarray}
\widetilde{G}(r_{ij})=\textbf{\emph{F}}_{i}\cdot\textbf{\emph{F}}_{j}\mathop{\sum}\limits_{k=1}^{3}\frac{2\alpha_{k}}{3\sqrt{\pi}r_{ij}}\int^{\tau_{k}r_{ij}}_{0}e^{-x^{2}}dx
\end{eqnarray}
and $\tau_{k}=\frac{1}{\sqrt{\frac{1}{\sigma_{ij}^{2}}+\frac{1}{\gamma_{k}^{2}}}}$.

In Eq.(10), $\widetilde{H}^{\mathrm{hyp}}$ is the color-hyperfine interaction which contains a tensor interaction and a contact interaction,
\begin{eqnarray}
\widetilde{H}^{\mathrm{hyp}}_{ij}=\widetilde{H}^{\mathrm{tensor}}_{ij}+\widetilde{H}_{ij}^{\mathrm{c}}
\end{eqnarray}
with
\begin{align}
&\widetilde{H}^{\mathrm{tensor}}_{ij}=-\Big(\frac{\textbf{S}_{i}\cdot \textbf{r}_{ij}\textbf{S}_{j}\cdot \textbf{r}_{ij}/r_{ij}^{2}-\frac{1}{3}\textbf{S}_{i}\cdot\textbf{S}_{j}}{m_{i}m_{j}}\Big)\notag \\
&\times\Big(\frac{\partial^{2}}{\partial r_{ij}^{2}}-\frac{1}{r_{ij}}\frac{\partial}{\partial r_{ij}}\Big)\widetilde{G}_{ij}^{\mathrm{t}},
\end{align}
\begin{eqnarray}
\widetilde{H}^{\mathrm{c}}_{ij}=\frac{2\textbf{S}_{i}\cdot\textbf{S}_{j}}{3m_{i}m_{j}}\bigtriangledown^{2}\widetilde{G}_{ij}^{\mathrm{c}}
\end{eqnarray}
For the spin-orbit interaction, it can also be divided into two parts,
\begin{eqnarray}
\widetilde{H}^{\mathrm{so}}_{ij}=\widetilde{H}^{\mathrm{so(v)}}_{ij}+\widetilde{H}_{ij}^{\mathrm{so(s)}},
\end{eqnarray}
with
\begin{align}
&\widetilde{H}^{\mathrm{so(v)}}_{ij}=\frac{\textbf{S}_{i}\cdot \textbf{L}_{ij}}{2m_{i}^{2}r_{ij}}\frac{\partial \widetilde{G}^{\mathrm{so(v)}}_{ii}}{\partial r_{ij}}+\frac{\textbf{S}_{j}\cdot \textbf{L}_{ij}}{2m_{j}^{2}r_{ij}}\frac{\partial \widetilde{G}^{\mathrm{so(v)}}_{jj}}{\partial r_{ij}}\notag \\
&+\frac{(\textbf{S}_{i}+\textbf{S}_{j})\cdot \textbf{L}_{ij}}{m_{i}m_{j}r_{ij}}\frac{1}{r_{ij}}\frac{\partial \widetilde{G}^{\mathrm{so(v)}}_{ij}}{\partial r_{ij}}
\end{align}
and
\begin{align}
\widetilde{H}^{\mathrm{so(s)}}_{ij}=-\frac{\textbf{S}_{i}\cdot \textbf{L}_{ij}}{2m_{i}^{2}r_{ij}}\frac{\partial \widetilde{S}^{\mathrm{so(s)}}_{ii}}{\partial r_{ij}}-\frac{\textbf{S}_{j}\cdot \textbf{L}_{ij}}{2m_{j}^{2}r_{ij}}\frac{\partial \widetilde{S}^{\mathrm{so(s)}}_{jj}}{\partial r_{ij}}
\end{align}
In Eqs.(18),(19),(21) and (22), $\widetilde{G}^{\mathrm{t}}_{ij}$, $\widetilde{G}^{\mathrm{c}}_{ij}$, $\widetilde{G}^{\mathrm{so(v)}}_{ij}$ and $\widetilde{S}^{\mathrm{so(s)}}_{ii}$ are achieved from $\widetilde{G}(r_{ij})$ and $\widetilde{S}(r_{ij})$ by introducing momentum-dependent factors,
\begin{align}
\widetilde{G}^{\mathrm{t}}_{ij}=\Big(\frac{m_{i}m_{j}}{E_{i}E_{j}}\Big)^{\frac{1}{2}+\epsilon_{\mathrm{t}}}\widetilde{G}(r_{ij})\Big(\frac{m_{i}m_{j}}{E_{i}E_{j}}\Big)^{\frac{1}{2}+\epsilon_{\mathrm{t}}}
\end{align}
\begin{align}
\widetilde{G}^{\mathrm{c}}_{ij}=\Big(\frac{m_{i}m_{j}}{E_{i}E_{j}}\Big)^{\frac{1}{2}+\epsilon_{\mathrm{c}}}\widetilde{G}(r_{ij})\Big(\frac{m_{i}m_{j}}{E_{i}E_{j}}\Big)^{\frac{1}{2}+\epsilon_{\mathrm{c}}}
\end{align}
\begin{align}
\widetilde{G}^{\mathrm{so(v)}}_{ij}=\Big(\frac{m_{i}m_{j}}{E_{i}E_{j}}\Big)^{\frac{1}{2}+\epsilon_{\mathrm{so(v)}}}\widetilde{G}(r_{ij})\Big(\frac{m_{i}m_{j}}{E_{i}E_{j}}\Big)^{\frac{1}{2}+\epsilon_{\mathrm{so(v)}}}
\end{align}
\begin{align}
\widetilde{S}^{\mathrm{so(s)}}_{ii}=\Big(\frac{m_{i}^{2}}{E_{i}^{2}}\Big)^{\frac{1}{2}+\epsilon_{\mathrm{so(s)}}}\widetilde{S}(r_{ij})\Big(\frac{m_{i}^{2}}{E_{i}^{2}}\Big)^{\frac{1}{2}+\epsilon_{\mathrm{so(s)}}}
\end{align}
with $E_{i}=\sqrt{m_{i}^{2}+p_{ij}^{2}}$, and $\epsilon_{\mathrm{t}}$, $\epsilon_{\mathrm{c}}$, $\epsilon_{\mathrm{so(v)}}$ and $\epsilon_{\mathrm{so(s)}}$ are free parameters which take the same values with those in Ref.\cite{GLY1}. The $p_{ij}$ is the magnitude of the momentum of either of the quarks in the $ij$ center-of-mass frame.

For a three-body system, the calculations of the Hamiltonian matrix elements become laborious even with Gaussian basis functions.
This process can be simplified by introducing the ISG basis functions. The Gaussian basis function of Eq. (3) is then substituted by the following ISG basis functions\cite{QM22},
\begin{align}
&\phi_{nlm_{l}}(\boldsymbol{r})=N_{nl}\lim_{\varepsilon\rightarrow 0}\frac{1}{(\nu_{n}\varepsilon)^{l}}\sum_{k=1}^{k_{max}}C_{lm_{l},k}e^{-\nu_{n}(\textbf{r}-\varepsilon \textbf{D}_{lm_{l},k})^{2}}
\end{align}
where $\varepsilon$ is the shifted distance of the Gaussian basis. Taking the limit $\varepsilon\rightarrow 0$ is to be carried out after the matrix elements have been calculated analytically. The coefficient $C_{lm_{l},k}$ and the shift-direction vector $\textbf{D}_{lm_{l},k}$ in Eq. (27) are dimensionless numbers independent of $\nu_{n}$ and $\varepsilon$, and they can be described as,
\begin{flalign}
C_{lm_{l},k}\equiv\sum_{j=0}^{\big[\frac{l-m_{l}}{2}\big]}A_{lm_{l},j}\sum_{s=0}^{p}\sum_{t=0}^{q}\sum_{u=0}^{j}(-1)^{l-u-t-s}\binom{p}{s}\binom{q}{t}\binom{j}{u}
\end{flalign}
where $p=l-m_{l}-2j$, $q=j+m_{l}$ and
\begin{flalign}
\textbf{D}_{lm_{l},k}\equiv\frac{1}{l}[(2s-p)\textbf{a}_{z}+(2t-q)\textbf{a}_{xy}+(2u-j)\textbf{a}_{xy}^{*}]
\end{flalign}
with
\begin{flalign}
A_{lm_{l},j}=\Big[\frac{(2l+1)(l-m_{l})!}{4\pi(l+m_{l})!}\Big]^{\frac{1}{2}}\frac{(l+m_{l})!(-1)^{j}}{(-2)^{m_{l}}4^{j}j!(m_{l}+j)!(l-m_{l}-2j)!}
\end{flalign}
In relation (29), $\textbf{a}_{z}$, $\textbf{a}_{xy}$ and $\textbf{a}_{xy}^{*}$ are called the shift vectors that are defined as $\textbf{a}_{z}\equiv(0,0,1)$, $\textbf{a}_{xy}\equiv(1,i,0)$, $\textbf{a}_{xy}^{*}\equiv(1,-i,0)$. The spherical harmonic in Eq. (3) is substituted by a sets of coefficients $C_{lm_{l}k}$ and vectors $\textbf{D}_{lm_{l},k}$. Thus, the tedious angular momentum algebra and angle integration are avoided, which makes the matrix element calculation easy in practice.
For more details about the calculations of the Hamiltonian matrix elements, one can consults our previous work\cite{GLY1}.

After all of the matrix elements are evaluated, the mass spectra can be obtained by solving the generalized eigenvalue problem,
\begin{align}
\sum_{j=1}^{n_{max}^{2}}\Big(H_{ij}-EN_{ij}\Big)C_{j}=0, \quad (i=1-n_{max}^{2})
\end{align}
Here, $H_{ij}$ denotes the matrix element in the total color-flavor-spin-spatial base, $E$ is the eigenvalue, $C_{j}$ stands for the corresponding eigenvector, and $N_{ij}$ is the overlap matrix elements of the Gaussian functions, which arises from the nonorthogonality of the bases and can be expressed as,
\begin{align}
\notag
&N_{ij}\equiv \langle\phi_{n_{\rho_{a}}l_{\rho_{a}}m_{l_{\rho_{a}}}}|
\phi_{n_{\rho_{b}}l_{\rho_{b}}m_{l_{\rho_{b}}}}\rangle \notag \times\langle\phi_{n_{\lambda_{a}}l_{\lambda_{a}}m_{l_{\lambda_{a}}}}
|\phi_{n_{\lambda_{b}}l_{\lambda_{b}}m_{l_{\lambda_{b}}}}\rangle \notag \\
&=\Big(\frac{2\sqrt{\nu_{n_{\rho_{a}}}\nu_{n_{\rho_{b}}}}}{\nu_{n_{\rho_{a}}}+\nu_{n_{\rho_{b}}}}\Big)^{l_{\rho_{a}}+3/2}\times
\Big(\frac{2\sqrt{\nu_{n_{\lambda_{a}}}\nu_{n_{\lambda_{b}}}}}{\nu_{n_{\lambda_{a}}}+\nu_{n_{\lambda_{b}}}}\Big)^{l_{\lambda_{a}}+3/2} &
\end{align}

\section{Numerical results and discussions}

\subsection{ Numerical stabilities and $\rho$-modes}
\begin{figure}[htbp]
\centering
\includegraphics[height=4.7cm,width=6.5cm]{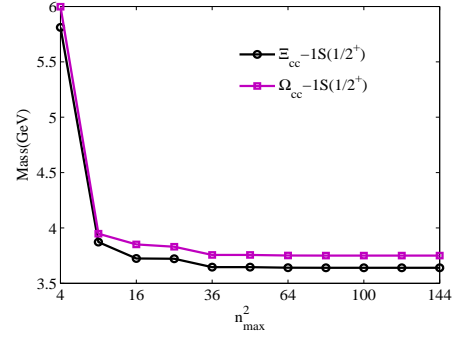}
\caption{Convergence of the energy of the lowest $\Xi_{cc}$ and $\Omega_{cc}$ for
increasing the number of bases functions.}
\label{figure0-1}
\end{figure}
\begin{figure}[htbp]
\centering
\includegraphics[height=4.7cm,width=6.5cm]{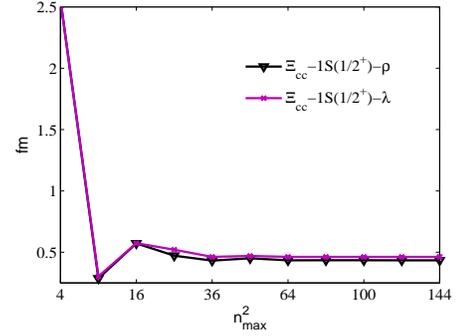}
\caption{Convergence of the r.m.s. radius $\sqrt{\langle r_{\rho}^{2}\rangle}$ and $\sqrt{\langle r_{\lambda}^{2}\rangle}$ of the lowest $\Xi_{cc}$ for
increasing the number of bases functions.}
\label{figure0-2}
\end{figure}
The parameters used in the Hamiltonian in Eq.(10) are the same as those in our previous work\cite{GLY1,ZYL1} where the experimental masses of single heavy baryons were well reproduced. In order to investigate the convergence and stability of the numerical results, we plot the masses of the lowest lying $\Xi_{cc}(\frac{1}{2}^{+})$ and $\Omega_{cc}(\frac{1}{2}^{+})$ baryons in Fig. \ref{figure0-1} and the r.m.s. radii $\sqrt {\langle {r_{\rho}^{2}}\rangle }$ and $\sqrt{\langle r_{\lambda}^{2}\rangle}$ of the lowest $\Xi_{cc}(\frac{1}{2}^{+})$ in Fig. \ref{figure0-2}. We can see that the results decrease with the basis number and converge to a stable value when the $n_{max}^{2}=100$. Thus, it is reliable for us to carry out the calculations with $100$ Gaussian bases in present work.
\begin{figure}[htbp]
\centering
\includegraphics[height=4.7cm,width=6.5cm]{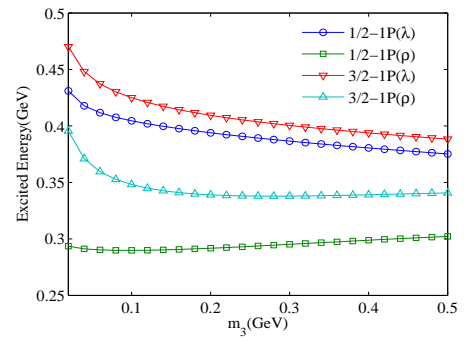}
\caption{Quark mass dependence of excited energy of doubly charmed baryons($\frac{1}{2}^{-}$,$\frac{3}{2}^{-}$) with $\lambda$-mode and $\rho$-mode}
\label{figure1-1}
\end{figure}
\begin{figure}[htbp]
\centering
\includegraphics[height=4.7cm,width=6.5cm]{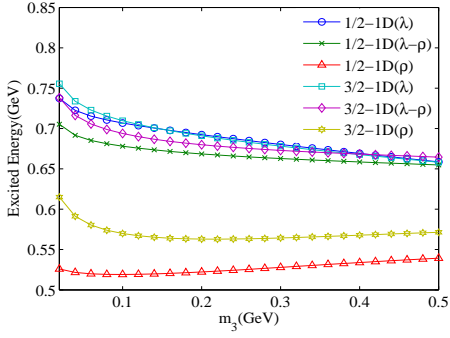}
\caption{Quark mass dependence of excited energy of doubly charmed baryons($\frac{1}{2}^{+}$,$\frac{3}{2}^{+}$) with $\lambda$-mode, $\rho$-mode and $\lambda$-$\rho$ mixing mode}
\label{figure1-2}
\end{figure}

For the orbital excitations of doubly charmed baryons, they can be classified by the orbital angular momentum $l_{\rho}$ and $l_{\lambda}$. For example, there are two orbital excitation modes $\lambda$- and $\rho$-mode with ($l_{\rho}$,$l_{\lambda}$)=($0$,$1$) and ($1$,$0$) for $P$-wave baryons. While there are three excitation modes for $D$-wave baryons with ($l_{\rho}$,$l_{\lambda}$)=($0$,$2$), ($2$,$0$) and ($1$,$1$), which are called the $\lambda$-mode, $\rho$-mode and $\lambda$-$\rho$ mixing mode, respectively. For higher orbital excited states, their situations are similar to $D$-wave baryons which also have three excitation modes. By changing the mass of light quark(denoted as $m_{3}$ in Figs. \ref{figure1-1}-\ref{figure1-2}) from $0.01\sim0.5$ GeV, we plot the excited energy of different excited modes in Figs. \ref{figure1-1}-\ref{figure1-2}. It is shown that the $\rho$-mode appears lower in excited energy than both the $\lambda$-mode and $\lambda$-$\rho$ mixing mode. This indicates the lowest states of doubly charmed baryons are dominated by the $\rho$-mode whether for $P$- or $D$-wave baryons. And this result is consistent with that of Ref.\cite{QM22}.

This above phenomenon originates from the interactions which are dependent on the orbital angular momentum(see Eqs. (21) and (22)). In these interactions, the energy is inversely proportional to the quark masses. Thus, the heavier the mass of quark, the lower the excited energy will be. For the doubly charmed baryons, the $\rho$-mode excitation between two charmed quarks is lower than the other two excited modes. The situation is just opposite for single heavy baryons that the $\lambda$-mode excitation between the center of mass of two light quarks and the heavy quark is the lowest. In the SU(3) limit in the light quark sector, these three orbitally excited modes will mix with each other.
\begin{table}[htbp]
\begin{ruledtabular}\caption{Masses(in MeV) for the ground states of $\Xi_{cc}$ and $\Omega_{cc}$ heavy baryons}
\label{tableI}
\begin{tabular}{c c c c c}
Baryons & $\Xi_{cc}(\frac{1}{2}^{+})$ &$\Xi_{cc}^{*}(\frac{3}{2}^{+})$ &$\Omega_{cc}(\frac{1}{2}^{+})$ & $\Omega_{cc}^{*}(\frac{3}{2}^{+})$ \\ \hline
Present work & 3640 & 3695 & 3750 &3799    \\
\cite{QM0} &3676 & 3753 &3815 & 3876  \\
\cite{QM4}&3620 & 3727 &3778 &3872   \\
\cite{QM2}& 3478 & 3610 &- & - \\
\cite{Bag1} &3520 & 3630 &3619 & 3721  \\
\cite{QM15} &3510 & 3548 &3719 & 3746  \\
\cite{Other4} & 3676& 3746 &3787 &3851 \\
\cite{QM14} &3613 & 3707 &3712 & 3795   \\
\cite{QM18} &3579 & 3708 &3718 & 3847 \\
\cite{QM19}& 3678 & 3752 &- &-  \\
\cite{Sum6}& - & 3690 &- &3780  \\
\cite{QM21}& 3532 & 3623 & 3667& 3758 \\
\cite{QM22}& 3685 & 3754 & 3832& 3883 \\
\cite{QM23} & 3606 & 3675 & 3715& 3772 \\
\cite{QM13} & 3612 & 3706 &3702 &3783 \\
\cite{BSE2} & 3547 & 3719 &3648 & 3770 \\
\cite{Bag2}& 3557 & 3661 &3710&3800 \\
\cite{QM25}   & 3520 & 3695 & 3650& 3810 \\
\cite{Bag4} & 3550 & 3590 &3730 & 3770 \\
\cite{QM27}& 3679 & 3763 & 3830& 3891 \\
\cite{Other1} & 3615 & 3747 & - & - \\
\cite{Other2}& 3627 & 3690 &3692 & 3756 \\
\cite{lattice1} & 3626 & 3693 &3719 & 3788 \\
\cite{Sum5}& 3630 & 3750 &3750 & 3850\\
\cite{BSE1}& 3620 & 3620 & 3720& 3720\\
\cite{Other3} & 3653 & 3741 & -& -\\
\cite{Bag3}& 3604 & 3714 & 3726& 3820 \\
\cite{QM34}  & 3620 & 3653 &3798 &3831 \\
\end{tabular}
\end{ruledtabular}
\end{table}

\subsection{ Mass spectra of $\Xi_{cc}$ and $\Omega_{cc}$}

For $\Xi_{cc}$ and $\Omega_{cc}$ baryons with $\rho$ excited mode, we obtain their r.m.s. radii and complete mass spectra with quantum numbers up to $n = 4$ and
$L = 4$.
Many collaborations have focused on the ground state masses of doubly charmed baryons, which results are listed in Table \ref{tableI} together with ours. From Table \ref{tableI}, we can see that our predicted mass for the ground state of $\Xi_{cc}$ is $3640$ MeV. Considering the model uncertainties, this result is consistent with the experimental data $3621.40$ MeV. In addition, our predictions for $\Xi_{cc}(\frac{1}{2}^{+},\frac{3}{2}^{+})$ and $\Omega_{cc}(\frac{1}{2}^{+},\frac{3}{2}^{+})$ are close to the results from Refs.\cite{QM4,QM13,QM14,Sum6,Bag3,lattice1,Other2}.
\begin{figure}[htbp]
\centering
\includegraphics[width=0.5\textwidth]{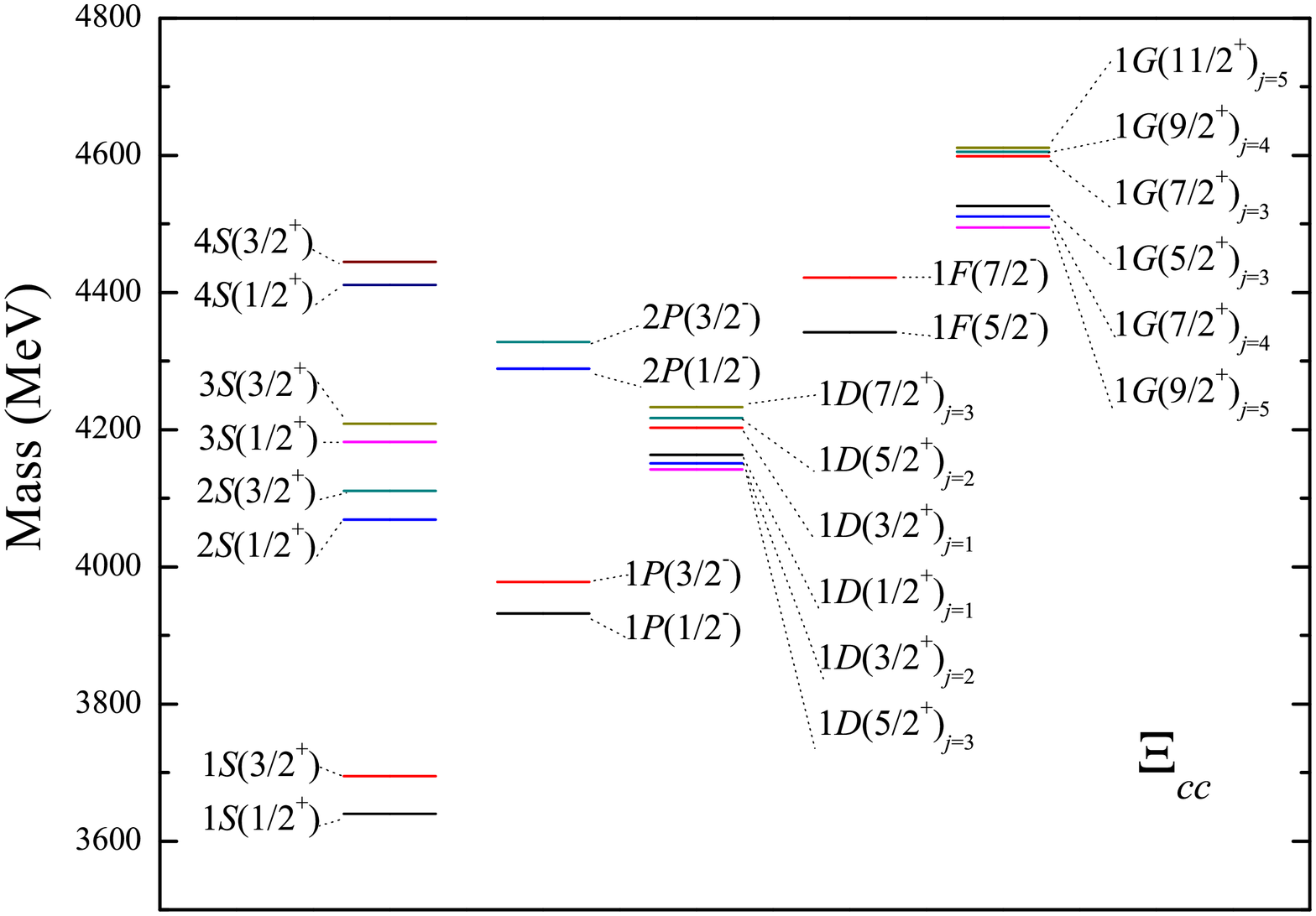}
\caption{Mass spectrum of $\Xi_{cc}$ family}
\label{figure1-3}
\end{figure}
\begin{figure}[htbp]
\centering
\includegraphics[width=0.5\textwidth]{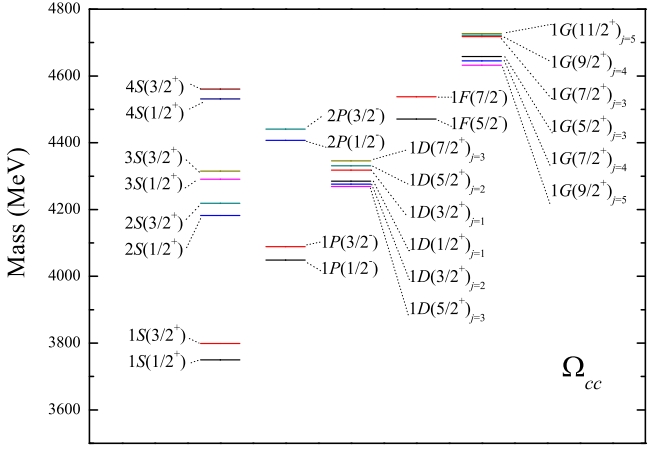}
\caption{Mass spectrum of $\Omega_{cc}$ family}
\label{figure1-4}
\end{figure}

For higher radial and orbital excitations together with the ground states, the predicted masses and the r.m.s. radii are shown in Tables \ref{tableIII}-\ref{tableIV} of Appendix A, where each state is characterized by the quantum numbers ($l_{\rho}$  $l_{\lambda}$ $L$ $s$ $j$) and $nL$($J^{P}$) in the first two columns. In order to see the feature of the mass spectra obviously, we show some of the results in Figs. \ref{figure1-3}-\ref{figure1-4}. From Tables \ref{tableIII}-\ref{tableIV} and Figs. \ref{figure1-3}-\ref{figure1-4}, we can see that the structure of the mass spectra of $\Xi_{cc}$ and $\Omega_{cc}$ are similar to each other. They have the following features: (1) The $1P$-wave doublet $(\frac{1}{2}^{-},\frac{3}{2}^{-})$ are the lowest excited states, and then is the $2S$ doublet $(\frac{1}{2}^{+},\frac{3}{2}^{+})$. This means they have good potentials to be observed in the future experiments. (2)After considering the spin-orbital interaction, there still exist degeneracy for higher orbital excited states with different $J^{P}$. For example, the excited states $1D(\frac{1}{2}^{+})_{j=1}$, $1D(\frac{3}{2}^{+})_{j=2}$ and $1D(\frac{5}{2}^{+})_{j=3}$ almost lie in the same energy level. (3)For the spin-doublet states, the energy of the $J = j+\frac{1}{2}$ state is higher than that of the $J = j-\frac{1}{2}$ state. (4)The energy difference between two adjacent radial excited states gradually decreases with radial quantum number $n$ increasing.

Because the results of this work cover 1S to 4D states, the open thresholds may occur in this large energy region. Under this circumstance, there are $qqq(q\overline{q})$ configurations possible in baryons, and these must have an effect on the constituent quark model. This coupled effect between bare three-quark state and meson-baryon state or between meson-baryon states can be studied by different methods\cite{couple1,couple2,couple3,coupleBS1,coupleBS2}.
Some light and single heavy baryons were already studied by considering this coupled-channel effect\cite{couple1,couple2,couple3,coupleBS1,coupleBS2}. It was indicated that this coupling has important influence on the light baryons and can significantly suppress the masses of these states\cite{couple1}. However, its influence on single heavy baryons may be limited\cite{couple1}. If this coupled-channel effect was also considered in the study of doubly charmed baryons, the masses of these states may also be suppressed, but limited. Nevertheless, its influence on doubly charmed baryons is an interesting problem worth studying and discussing in the future.
\subsection{ The r.m.s. radii and radial density distributions}
\begin{figure}[htbp]
  \centering
   \subfigure[]{
   \begin{minipage}{3.9cm}
   \centering
   \includegraphics[width=4.5cm]{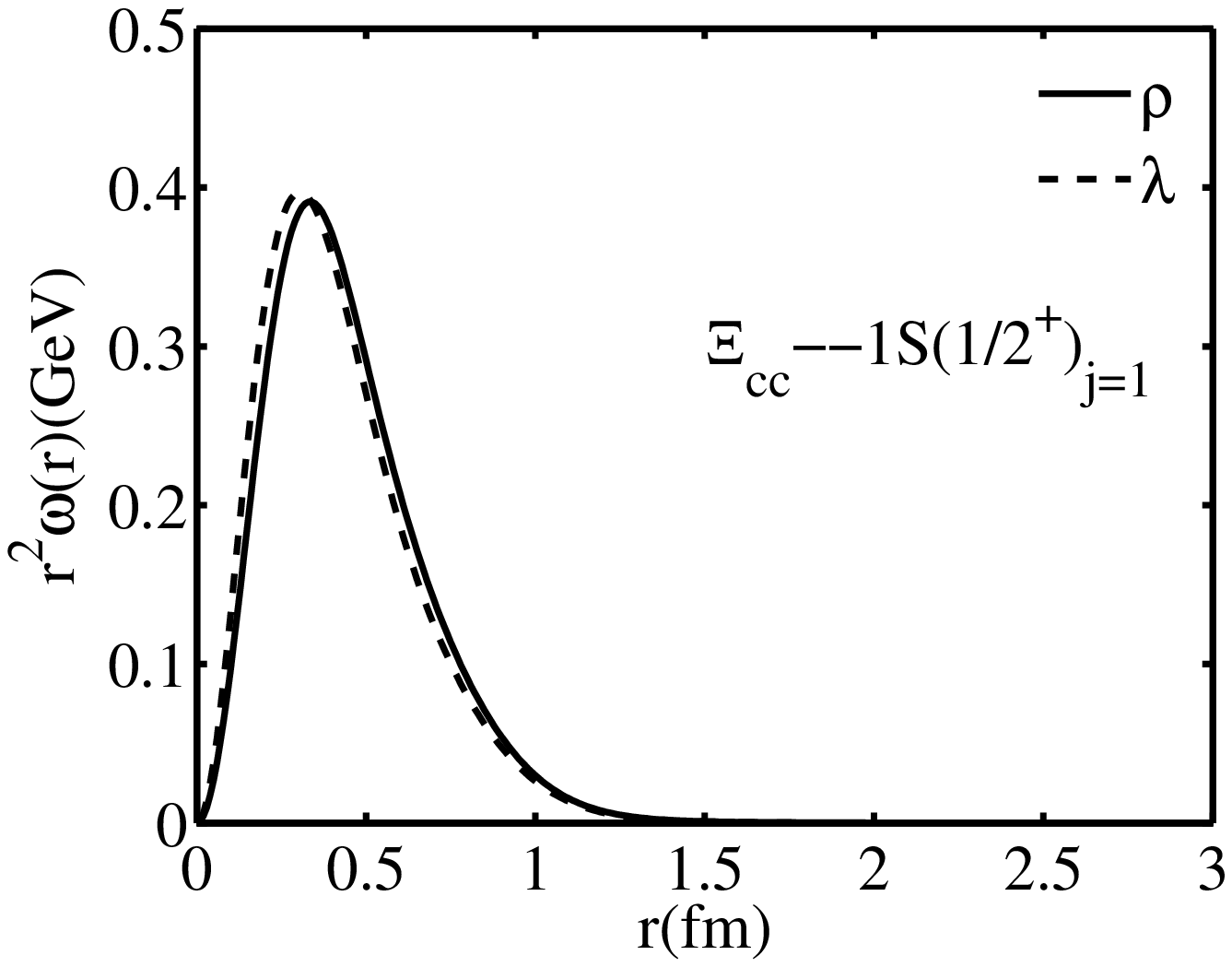}
  \end{minipage}
  }
 \subfigure[]{
   \begin{minipage}{3.9cm}
   \centering
   \includegraphics[width=4.5cm]{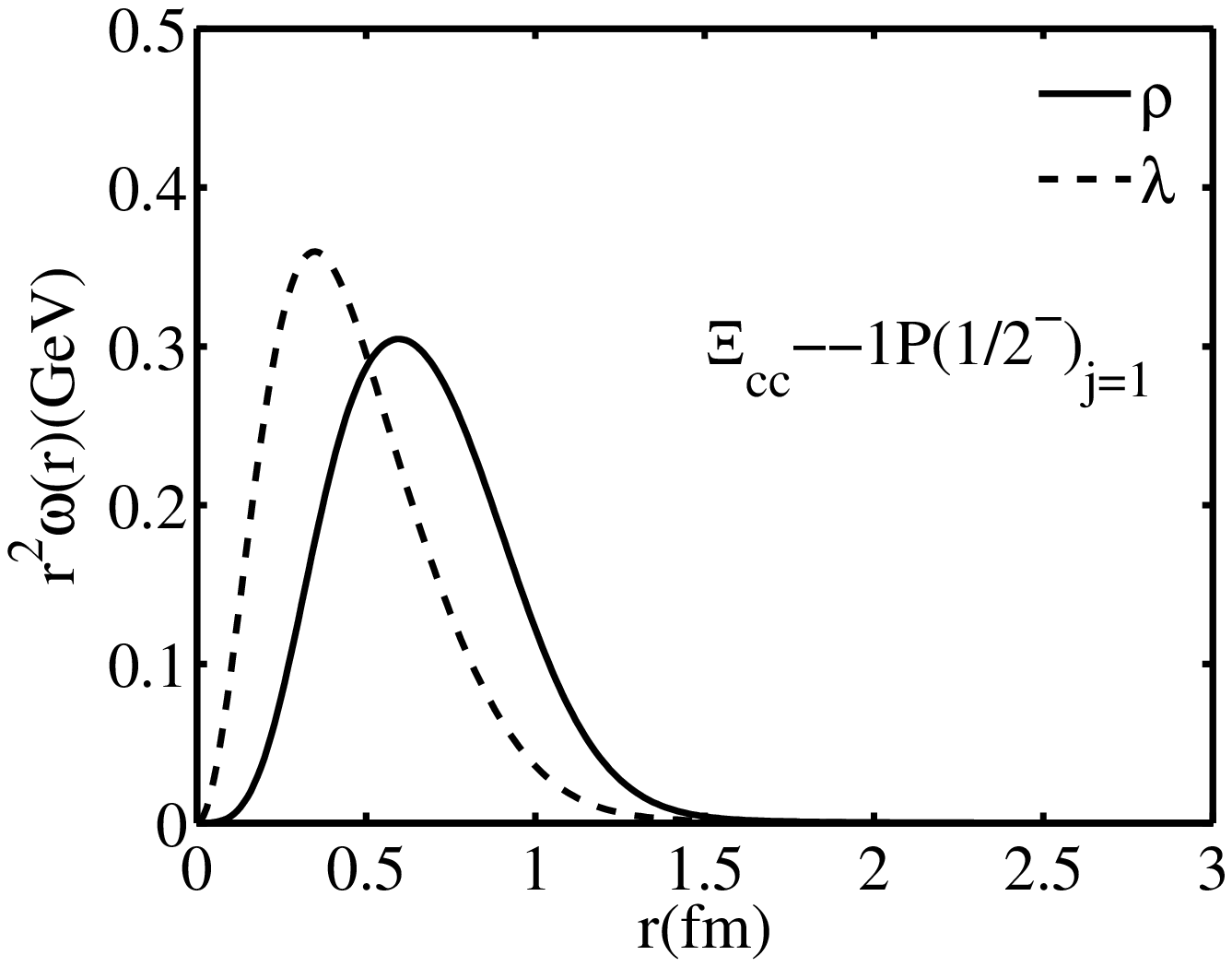}
  \end{minipage}
  }
   \subfigure[]{
   \begin{minipage}{3.9cm}
   \centering
   \includegraphics[width=4.5cm]{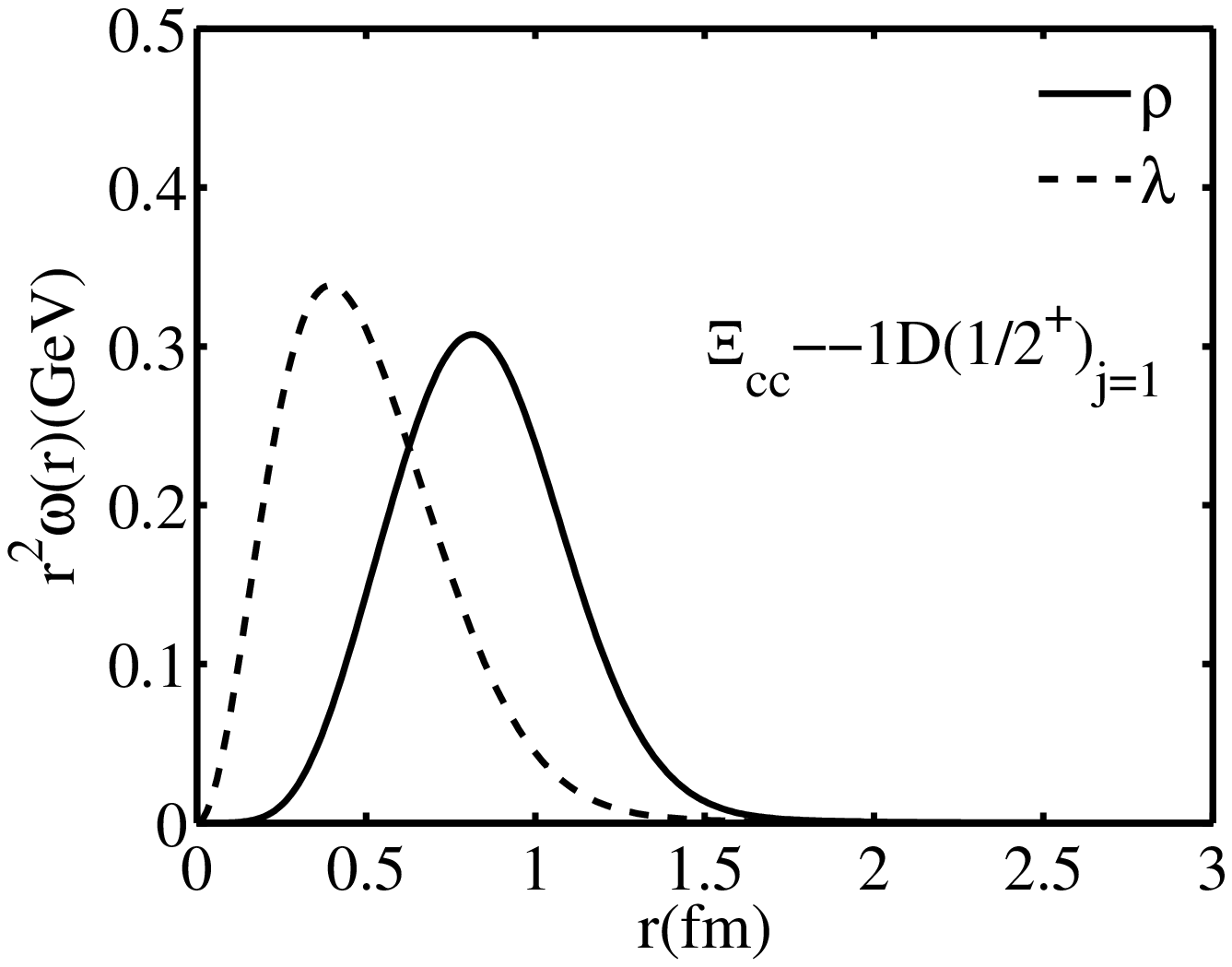}
  \end{minipage}
  }
    \subfigure[]{
   \begin{minipage}{3.9cm}
   \centering
   \includegraphics[width=4.5cm]{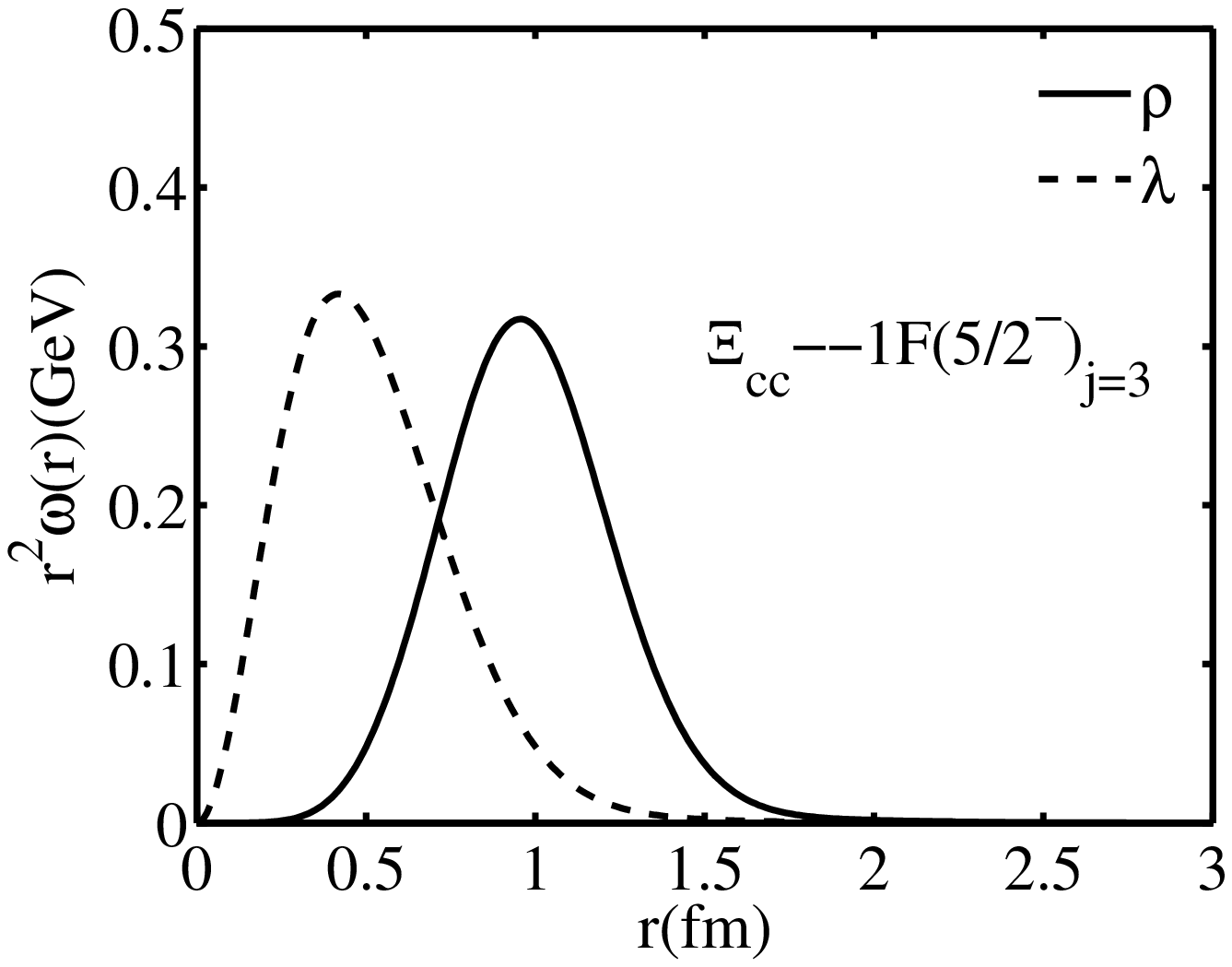}
  \end{minipage}
  }
  \caption{Radial density distributions for some $1S\sim1F$ states in the $\Xi_{cc}$ family}
\label{figure2}
\end{figure}
\begin{figure}[htbp]
  \centering
   \subfigure[]{
   \begin{minipage}{3.9cm}
   \centering
   \includegraphics[width=4.5cm]{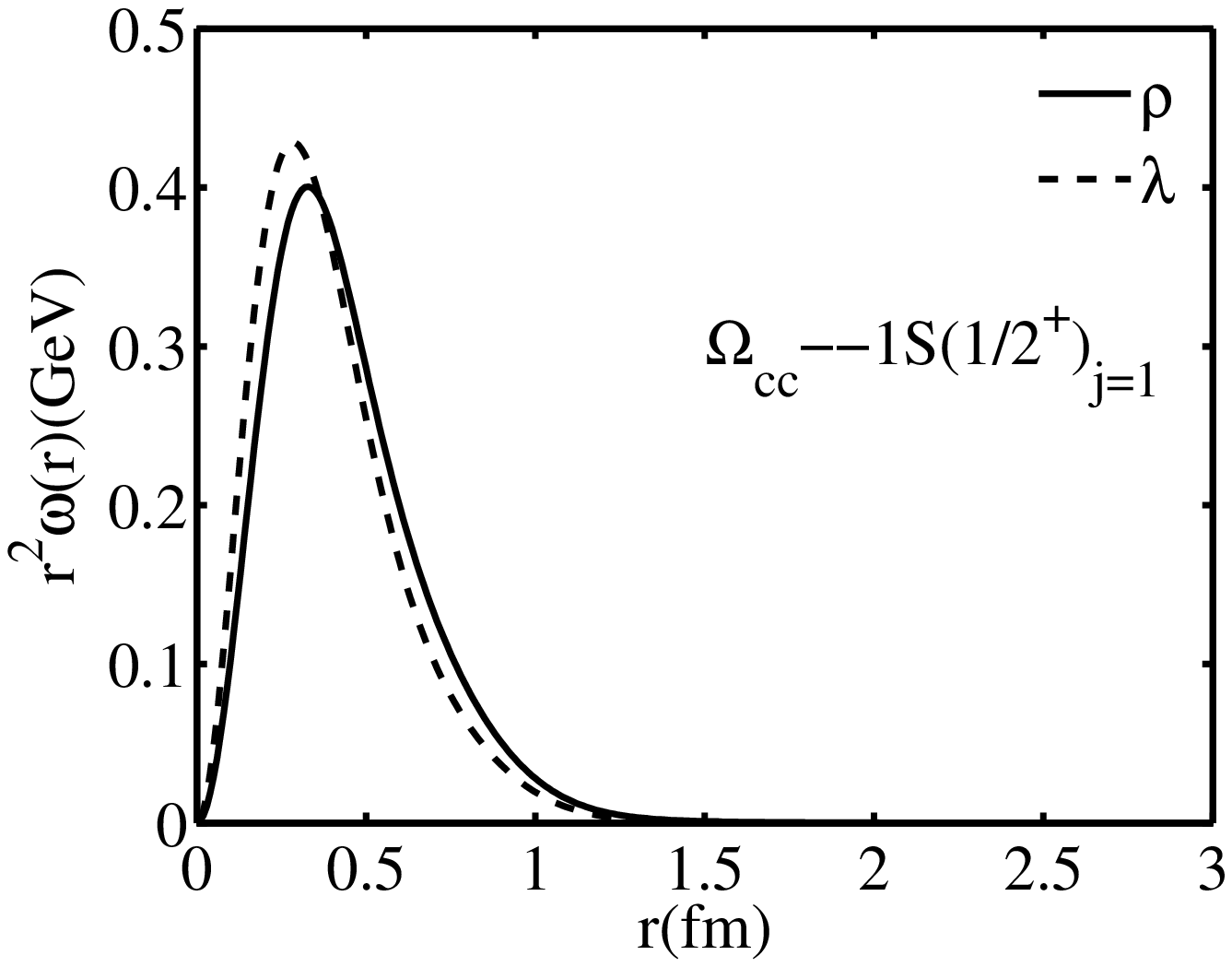}
  \end{minipage}
  }
 \subfigure[]{
   \begin{minipage}{3.9cm}
   \centering
   \includegraphics[width=4.5cm]{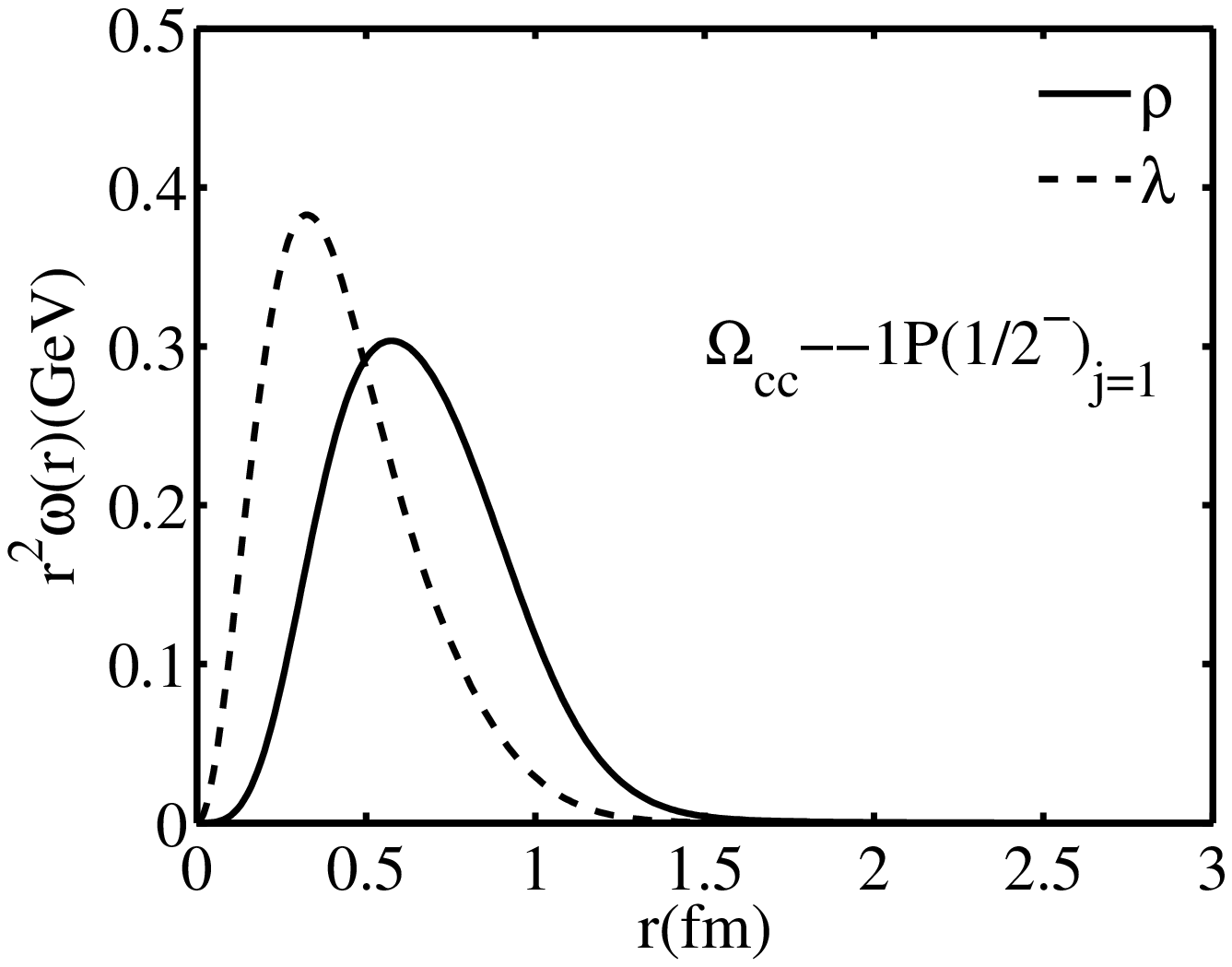}
  \end{minipage}
  }
   \subfigure[]{
   \begin{minipage}{3.9cm}
   \centering
   \includegraphics[width=4.5cm]{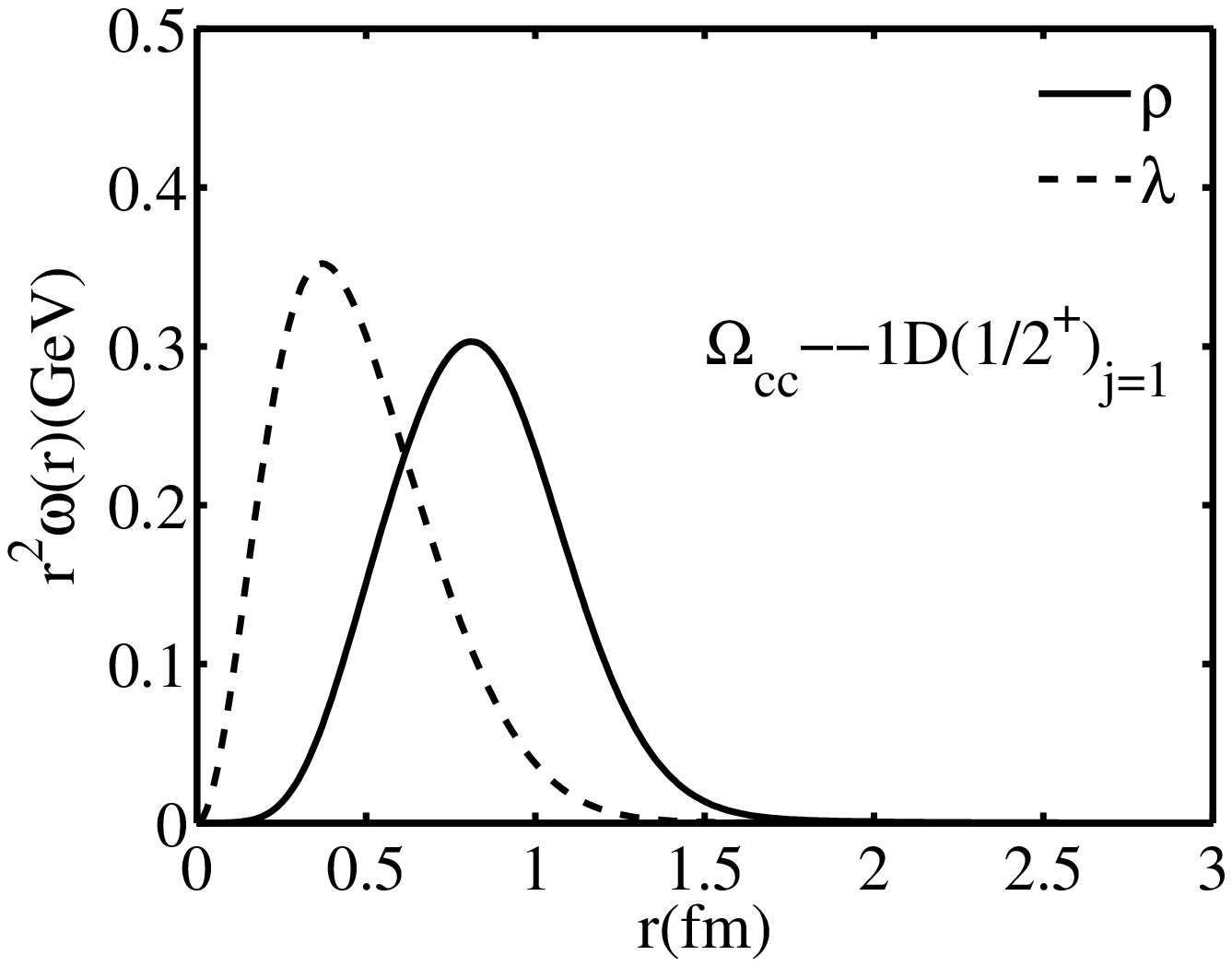}
  \end{minipage}
  }
    \subfigure[]{
   \begin{minipage}{3.9cm}
   \centering
   \includegraphics[width=4.5cm]{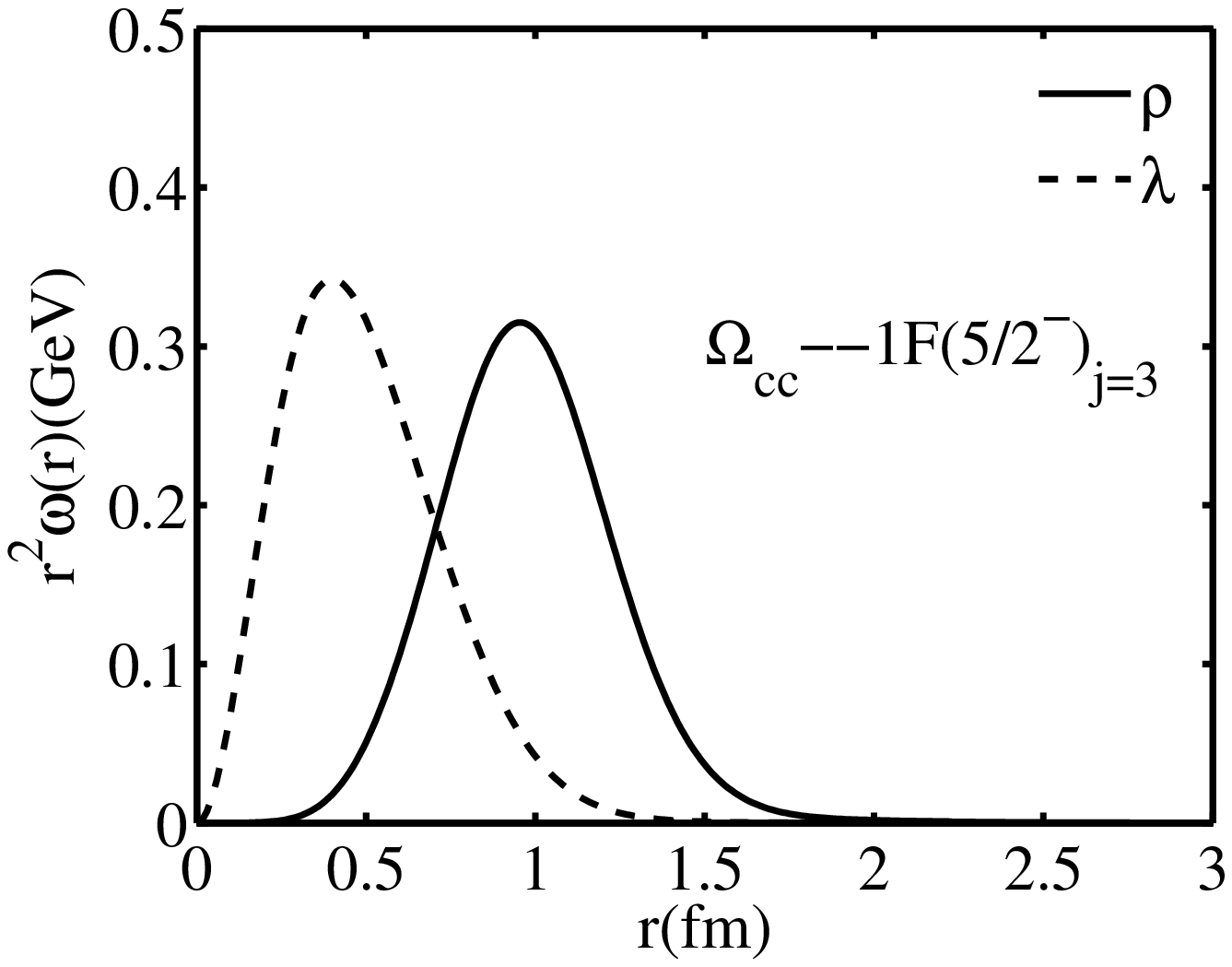}
  \end{minipage}
  }
  \caption{Radial density distributions for some $1S\sim1F$ states in the $\Omega_{cc}$ family}
  \label{figure3}
\end{figure}
\begin{figure}[htbp]
  \centering
   \subfigure[]{
   \begin{minipage}{3.9cm}
   \centering
   \includegraphics[width=4.5cm]{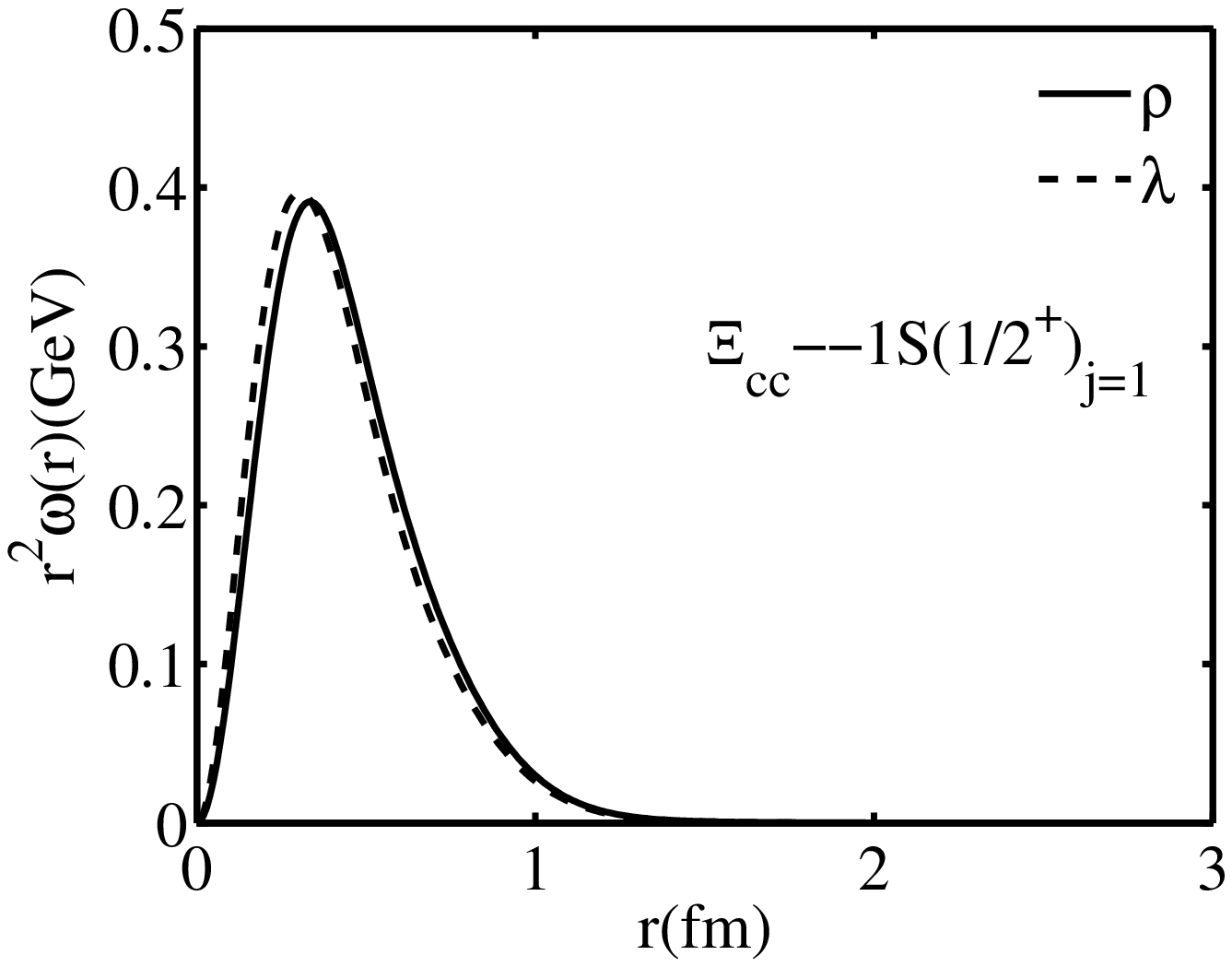}
  \end{minipage}
  }
 \subfigure[]{
   \begin{minipage}{3.9cm}
   \centering
   \includegraphics[width=4.5cm]{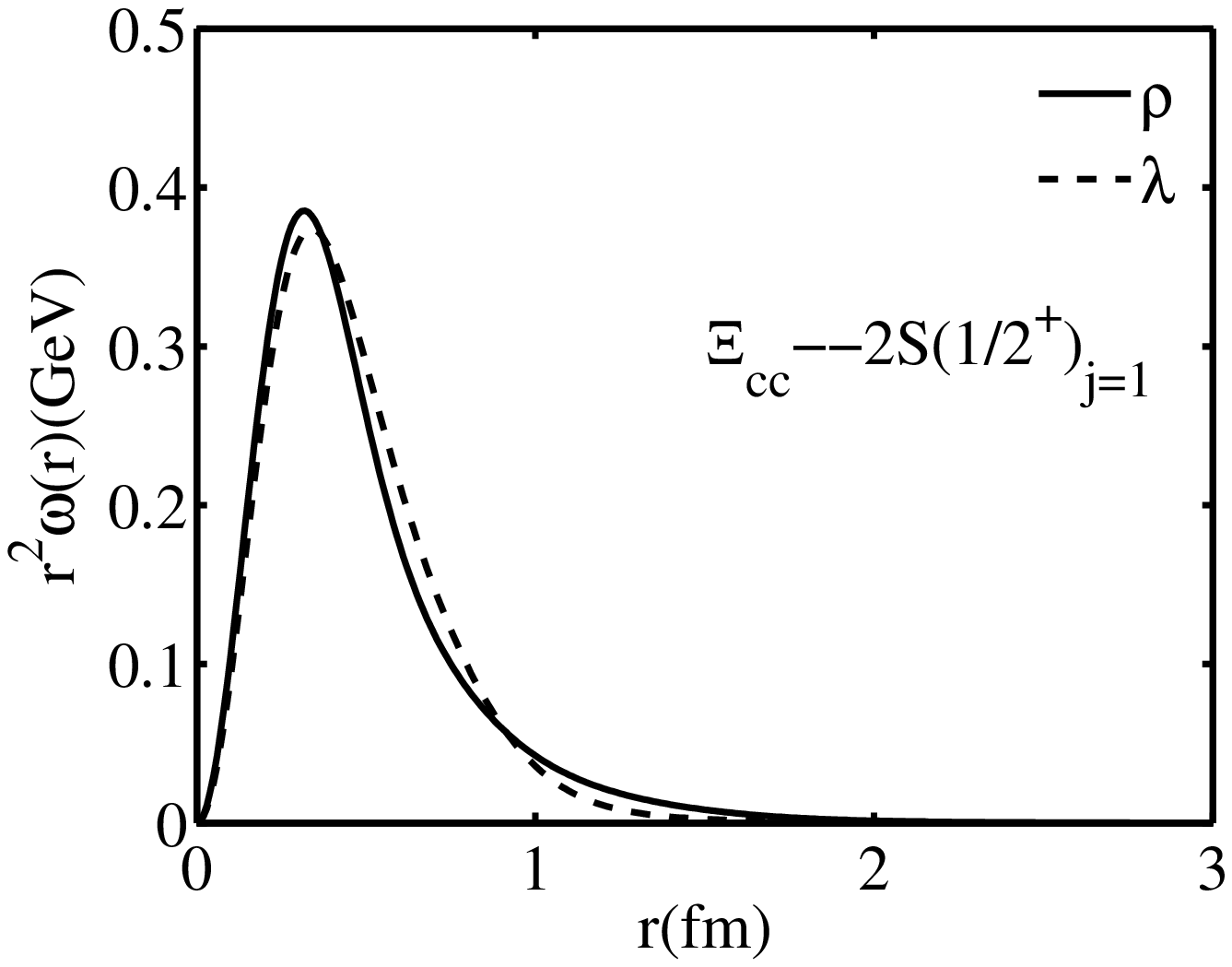}
  \end{minipage}
  }
   \subfigure[]{
   \begin{minipage}{3.9cm}
   \centering
   \includegraphics[width=4.5cm]{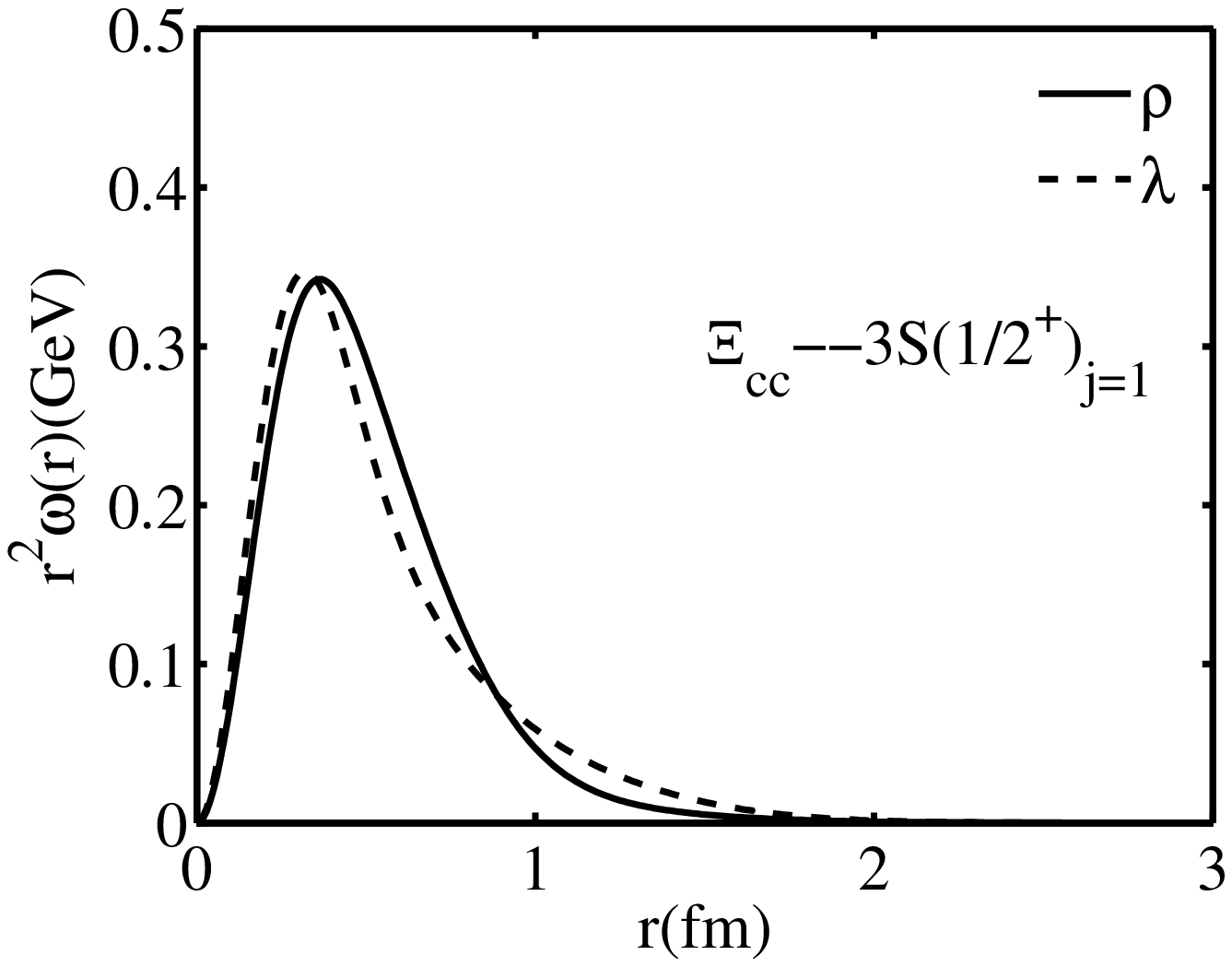}
  \end{minipage}
  }
   \subfigure[]{
   \begin{minipage}{3.9cm}
   \centering
   \includegraphics[width=4.5cm]{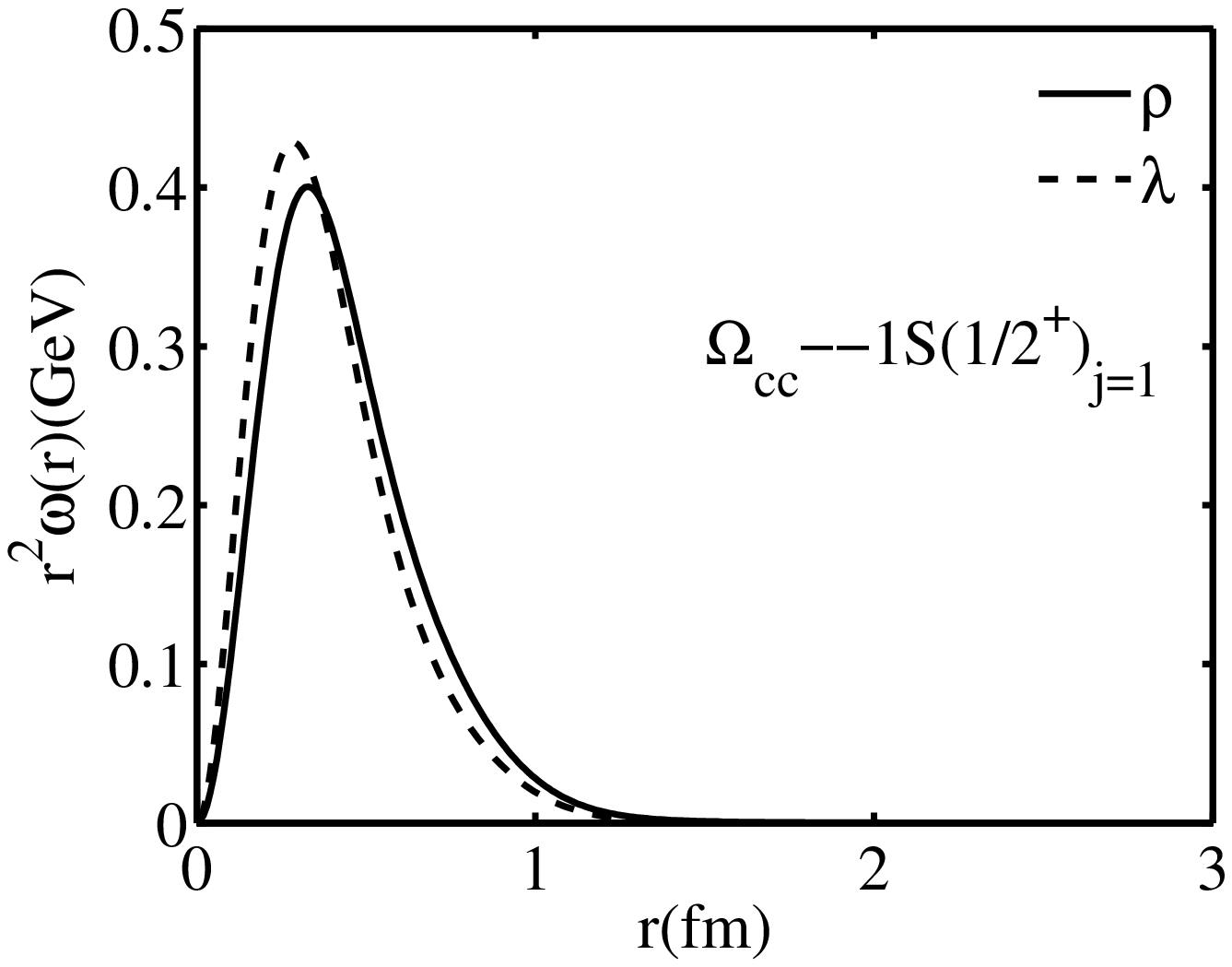}
  \end{minipage}
  }
 \subfigure[]{
   \begin{minipage}{3.9cm}
   \centering
   \includegraphics[width=4.5cm]{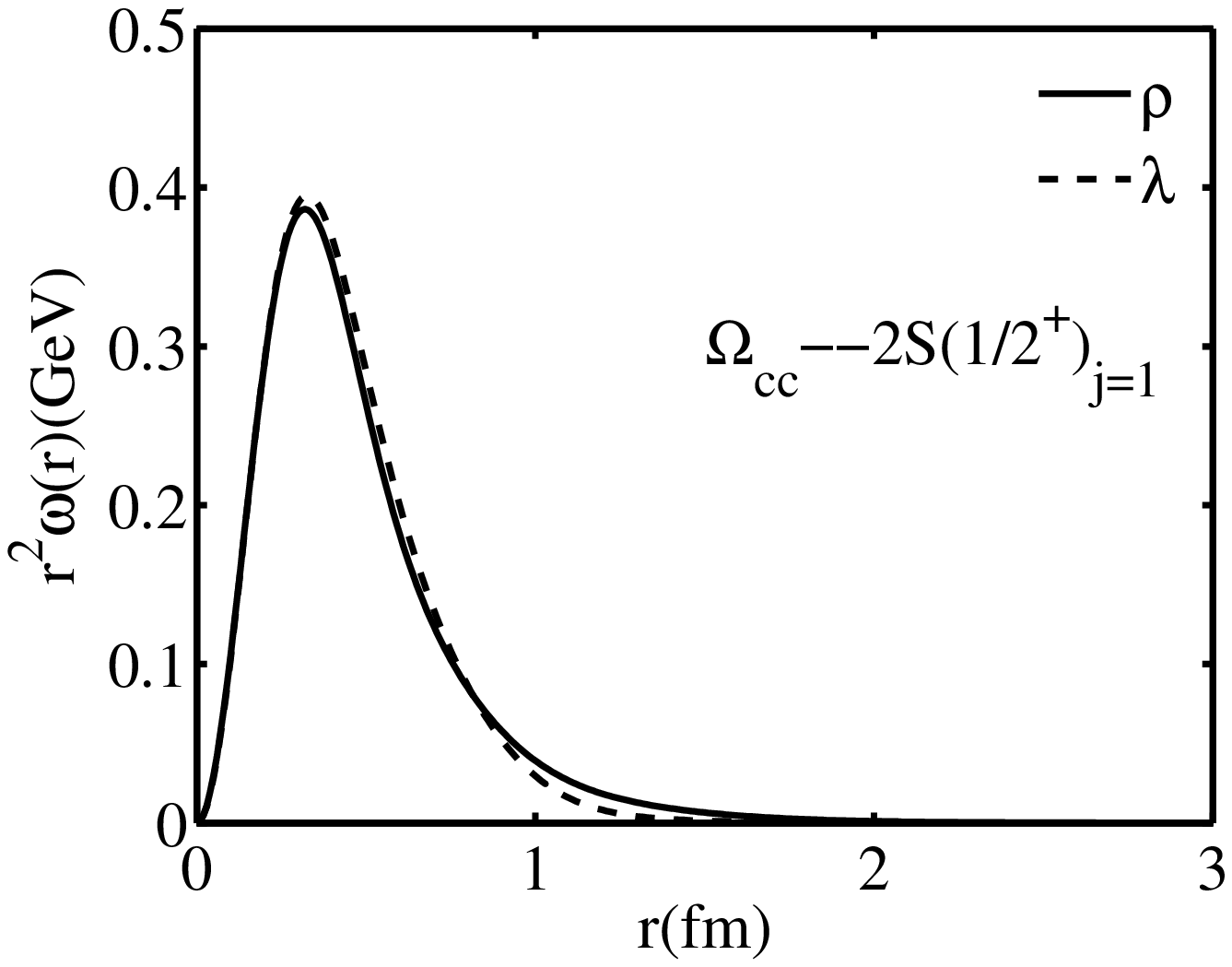}
  \end{minipage}
  }
   \subfigure[]{
   \begin{minipage}{3.9cm}
   \centering
   \includegraphics[width=4.5cm]{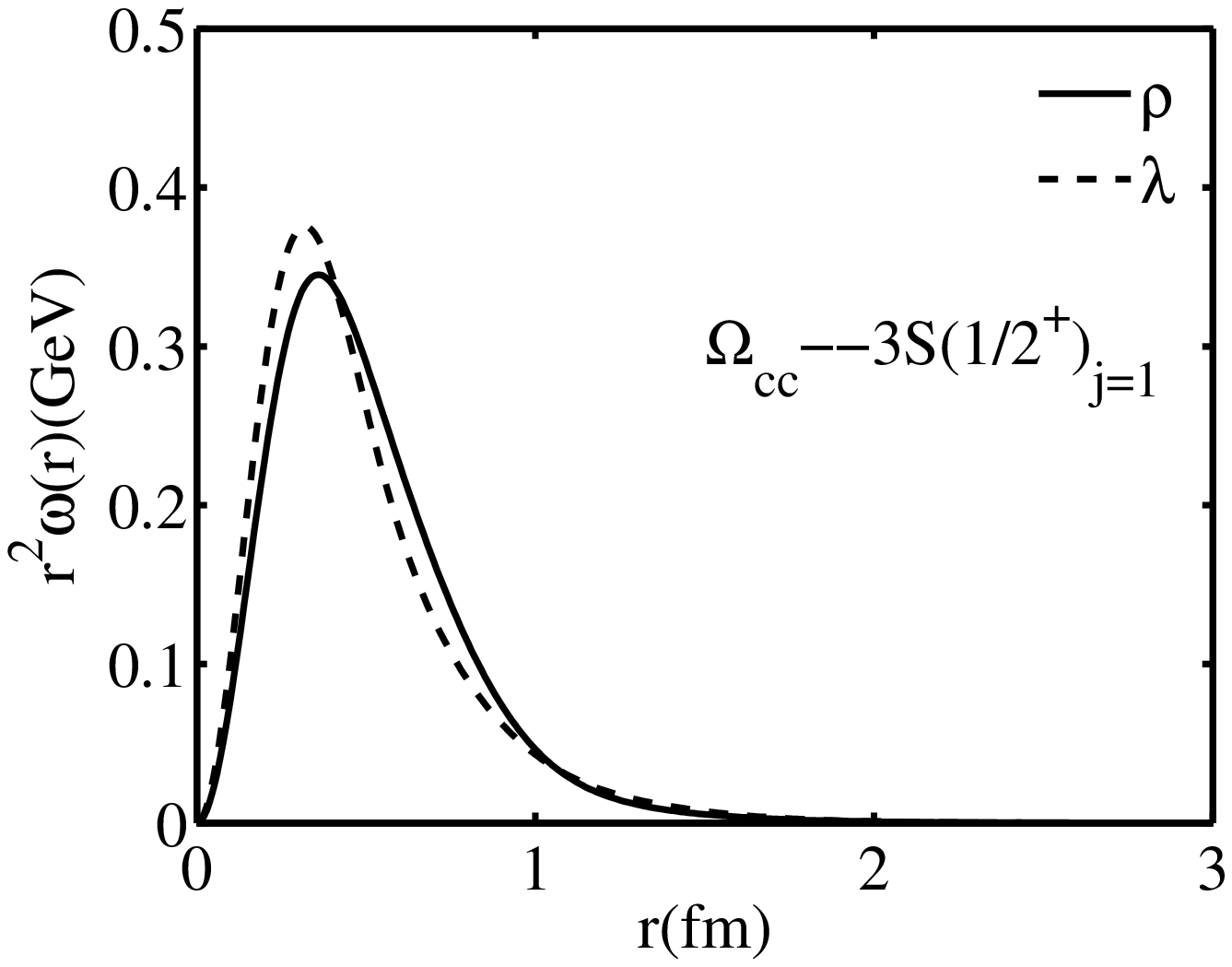}
  \end{minipage}
  }
  \caption{Radial density distributions for $1S\sim3S$ states in the $\Xi_{cc}$ family(a-c) and the $\Omega_{cc}$ family(d-f).}
\label{figure4}
\end{figure}
The r.m.s. radius and radial density distribution of the baryons are important for testing various conjectures about strongly interacting systems. Thus, using the wave functions obtained from quark model, we also study these parameters.  The r.m.s. radii for doubly charmed states are also shown in Tables \ref{tableIII}-\ref{tableIV}, and the radial density distributions are defined as,
\begin{eqnarray}
\notag
\omega(r_{\rho})=\int|\Psi(\textbf{r}_{\rho},\textbf{r}_{\lambda})|^{2}d\textbf{r}_{\lambda}d\Omega_{\rho} \\
\omega(r_{\lambda})=\int|\Psi(\textbf{r}_{\rho},\textbf{r}_{\lambda})|^{2}d\textbf{r}_{\rho}d\Omega_{\lambda}
\end{eqnarray}
where $\Omega_{\rho}$ and $\Omega_{\lambda}$ are the solid angles spanned by vectors $\textbf{r}_{\rho}$ and $\textbf{r}_{\lambda}$, respectively. Some of the results about the radial density distributions of baryons $\Xi_{cc}$ and $\Omega_{cc}$ are shown in Figs. \ref{figure2}-\ref{figure4}.

From Tables \ref{tableIII}-\ref{tableIV}, we can see the r.m.s. radii $\sqrt{\langle r_{\rho}^{2}\rangle}$ and $ \sqrt{\langle r_{\lambda}^{2}\rangle}$ of the $1S(\frac{1}{2}^{+})$ state are 0.435 fm, 0.462 fm for $\Xi_{cc}$, and 0.426 fm, 0.427 fm for $\Omega_{cc}$. In Refs.\cite{CM1,CM2,CM3}, their predicted r.m.s. radii for the ground state of $c\overline{c}$ meson are 0.449 fm, 0.445 fm, and 0.484 fm. These values are comparable with our results for $\sqrt{\langle r_{\rho}^{2}\rangle}$ of doubly charmed baryons. For the states with same radial quantum number $n$, the $\sqrt{\langle r_{\rho}^{2}\rangle}$ becomes larger obviously when the orbital angular momentum $L$ increases. However, $ \sqrt{\langle r_{\lambda}^{2}\rangle}$ increases a little with $L$ increasing. This is consistent with the results shown in Figs. \ref{figure2}-\ref{figure3}, where the peak of $r^{2}\omega(r_{\rho})$ shifts outward with $L$ increment and $r^{2}\omega(r_{\lambda})$ changes little. For the states with same angular momentum $L$, both $\sqrt{\langle r_{\rho}^{2}\rangle}$ and $\sqrt{\langle r_{\lambda}^{2}\rangle}$ increase with radial quantum number $n$. We can also see this feature from Fig. \ref{figure4}. It is shown that the peak of radial density distribution becomes lower from $1S\sim3S$ states and the peak position shifts outward slightly. Theoretically, the larger the r.m.s. radii become, the looser the baryons will be. We hope these results can help us to estimate the upper limit of the
mass spectra and to search for the new doubly charmed baryons in forthcoming experiments.

From Tables \ref{tableIII}-\ref{tableIV}, we can clearly see another interesting phenomenon that the order of r.m.s radii $\sqrt{\langle r_{\rho}^{2}\rangle}$ and $ \sqrt{\langle r_{\lambda}^{2}\rangle}$ behaves differently with the increment of the radial quantum numbers. For the states from $2D(\frac{1}{2}^{+})$ to $4D(\frac{1}{2}^{+})$ as examples, their r.m.s radii $\sqrt{\langle r_{\rho}^{2}\rangle}$ and $ \sqrt{\langle r_{\lambda}^{2}\rangle}$ alternately increase and decrease, if $\sqrt{\langle r_{\rho}^{2}\rangle}$ increase, $ \sqrt{\langle r_{\lambda}^{2}\rangle}$ will decrease. Because all of these D-wave states have definite orbital excitations with ($l_{\rho}$,$l_{\lambda}$)=(2,0), this phenomenon should be unrelated to orbital excitations. It may be related to the radial excitations, which can be denoted by ($n_{\rho}$,$n_{\lambda}$). The meaning of $n_{\rho}$ and $n_{\lambda}$ in Eq. (6) are similar to $l_{\rho}$ and $l_{\lambda}$ except that the former denotes the radial excitation and the latter is the orbital excitation. The different behaviors of the r.m.s. radii $\sqrt{\langle r_{\rho}^{2}\rangle}$ and $ \sqrt{\langle r_{\lambda}^{2}\rangle}$ reflect the distribution of radial excited energy between $n_{\rho}$ and $n_{\lambda}$. If the radial excited energy of $\rho$ mode is higher than that of $\lambda$ mode, then $\sqrt{\langle r_{\rho}^{2}\rangle}$ is larger than $ \sqrt{\langle r_{\lambda}^{2}\rangle}$. The radial excited energies distribute alternatively between $\rho$ and $\lambda$ mode with the increment of total radial quantum numbers. Thus, the phenomenon that we see in Tables \ref{tableIII}-\ref{tableIV} emerges.
\subsection{Regge trajectories of doubly charmed baryons $\Xi_{cc}$ and $\Omega_{cc}$}
\begin{figure}[htbp]
  \centering
   \subfigure[]{
   \begin{minipage}{3.9cm}
   \centering
   \includegraphics[width=4.5cm]{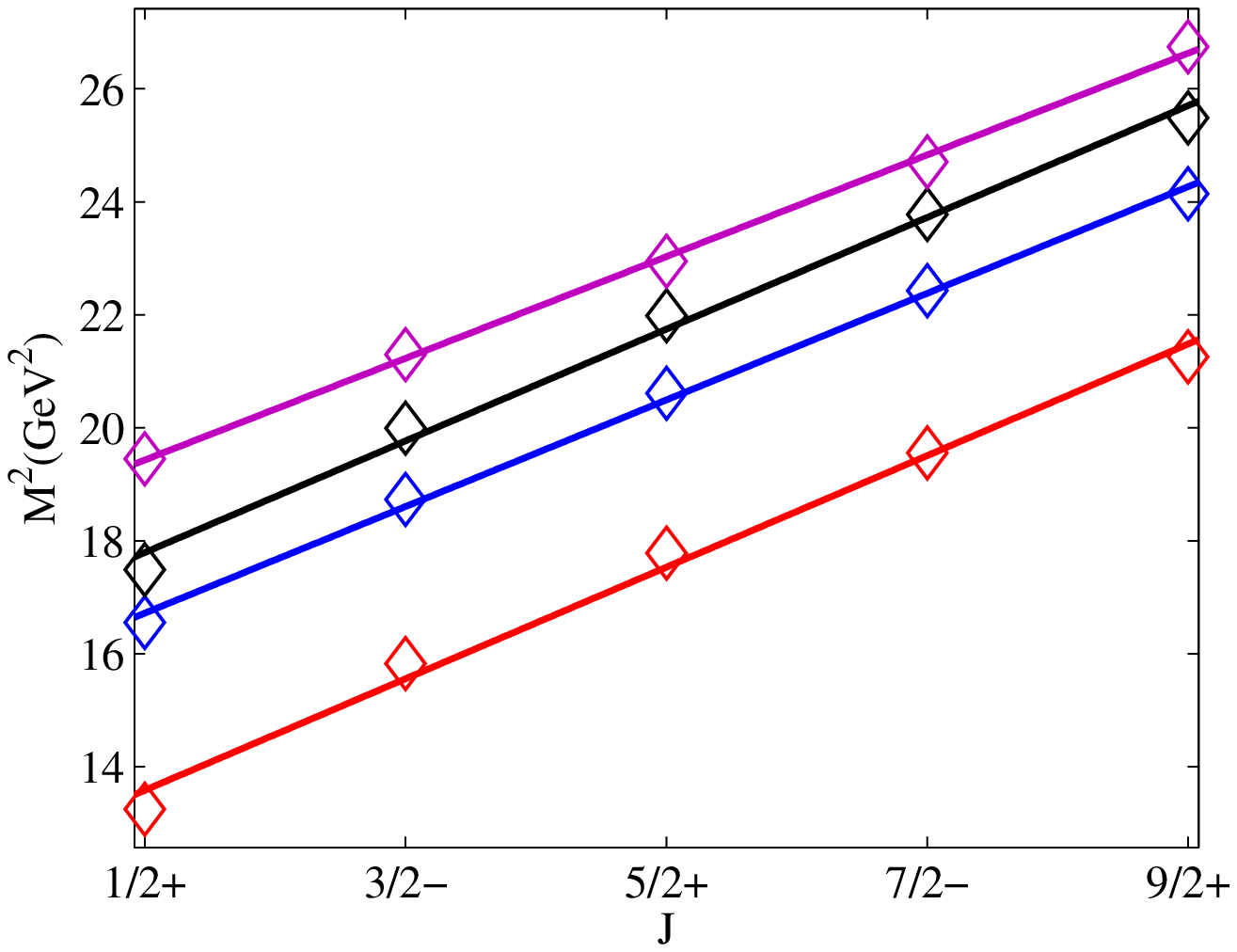}
  \end{minipage}
  }
 \subfigure[]{
   \begin{minipage}{3.9cm}
   \centering
   \includegraphics[width=4.5cm]{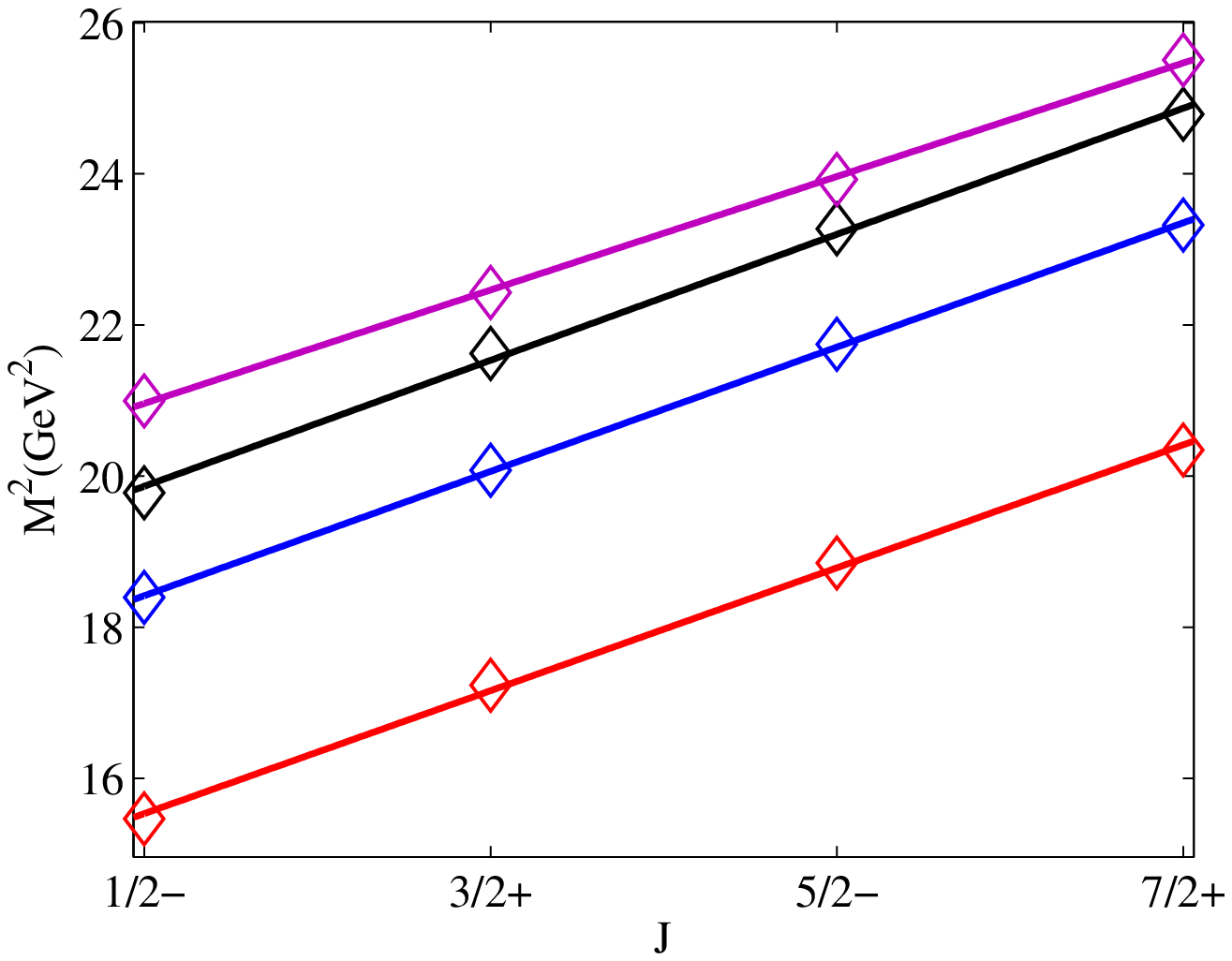}
  \end{minipage}
  }
   \subfigure[]{
   \begin{minipage}{3.9cm}
   \centering
   \includegraphics[width=4.5cm]{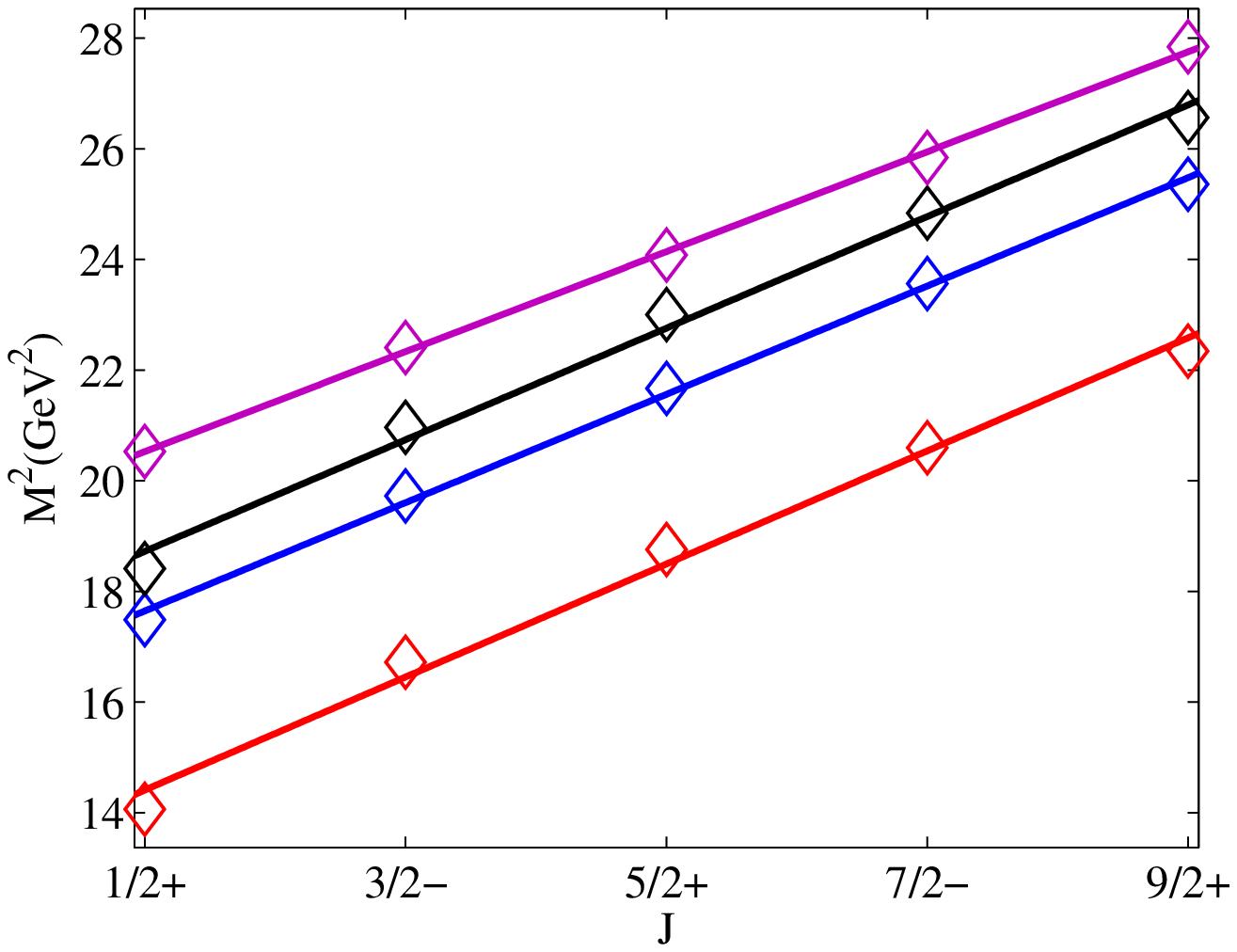}
  \end{minipage}
  }
    \subfigure[]{
   \begin{minipage}{3.9cm}
   \centering
   \includegraphics[width=4.5cm]{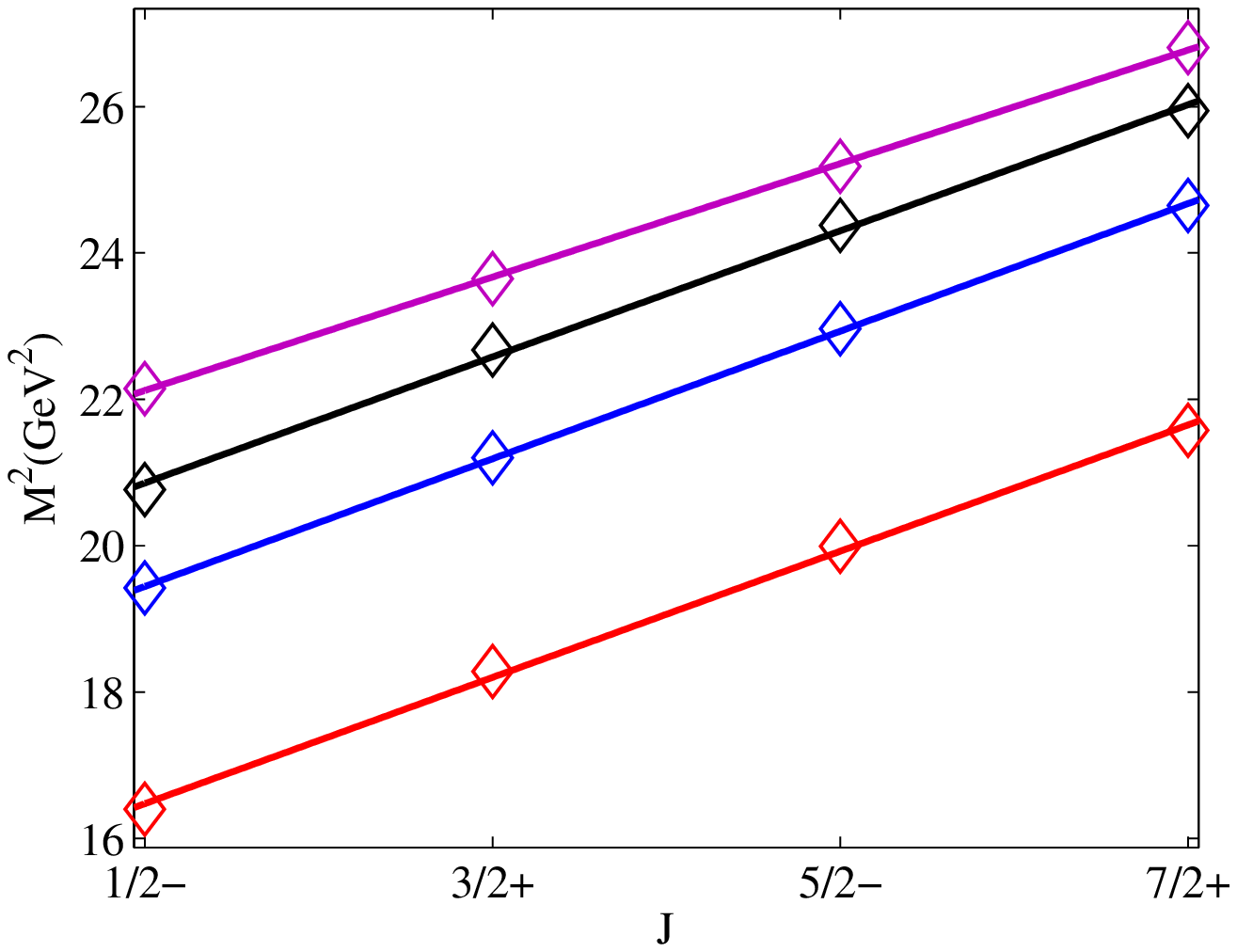}
  \end{minipage}
  }
  \caption{Parent and daughter Regge trajectories for the $\Xi_{cc}$ baryons with natural parity(a), unnatural parity(b) and $\Omega_{cc}$ baryons with natural parity(c), unnatural parity(d)}
\label{figure6}
\end{figure}
The Regge theory is very successful in studying the strong interaction at high energy and it is an indispensable tool in phenomenological studies for hadrons\cite{Regge1,Regge2,Regge3,Regge4,Regge5,Regge6,Regge7,Regge8,Regge9,Regge10}.
In our previous work, we have successfully constructed the Regge trajectories for the single heavy baryons\cite{GLY1,ZYL1}.
In the present work, we have obtained the $1S\sim4S$, $1P\sim4P$, $1D\sim4D$, $1F\sim4F$ and $1G\sim4G$ state masses for doubly charmed baryons. This makes it easy for us to construct their Regge trajectories in ($J$,$M^{2}$) plane.
The doubly charmed baryons are classified into two groups which have natural parity $S(\frac{1}{2}^{+})_{j=1}$, $P(\frac{3}{2}^{-})_{j=1}$, $D(\frac{5}{2}^{+})_{j=2}$, $F(\frac{7}{2}^{-})_{j=3}$, $G(\frac{9}{2}^{+})_{j=4}$ and unnatural parity $P(\frac{1}{2}^{-})_{j=1}$, $D(\frac{3}{2}^{+})_{j=2}$, $F(\frac{5}{2}^{-})_{j=3}$, $G(\frac{7}{2}^{+})_{j=4}$\cite{Ebert}. The Regge trajectories are presented in Fig. \ref{figure6}(a)-(b) for $\Xi_{cc}$ and in Fig. \ref{figure6}(c)-(d) for $\Omega_{cc}$, where the predicted masses in quark model are denoted by diamonds. The ground and radial excited states are plotted from bottom to top. We use the following definition about the ($J$,$M^{2}$) Regge trajectories,
\begin{eqnarray}
M^{2}=\alpha J+\alpha_{0}
\end{eqnarray}
\begin{table*}[htbp]
\begin{ruledtabular}\caption{Fitted parameters $\alpha$ and $\alpha_{0}$ for the slope and intercept of the ($J$,$M^2$) parent and daughter Regge trajectories for $\Xi_{cc}$ and $\Omega_{cc}$.}
\label{tableII}
\begin{tabular}{c c c| c c}
Trajectory&$\alpha$(Gev$^{2}$)&$\alpha_{0}$(Gev$^{2}$) &$\alpha$(Gev$^{2}$)&$\alpha_{0}$(Gev$^{2}$) \\ \hline
&$\Xi_{cc}(\frac{1}{2}^{+})$&~&$\Xi_{cc}(\frac{1}{2}^{-})$&~\\
parent&1.975$\pm$0.312&12.601$\pm$1.005&1.627$\pm$0.192&14.725$\pm$0.431\\
1 daughter&1.887$\pm$0.167&15.780$\pm$0.420&1.646$\pm$0.076&17.591$\pm$0.175\\
2 daughter&1.977$\pm$0.297&16.821$\pm$0.912&1.668$\pm$0.223&19.036$\pm$0.512\\
3 daughter&1.810$\pm$0.117&18.530$\pm$0.331&1.502$\pm$0.101&20.217$\pm$0.223\\ \hline
&$\Omega_{cc}(\frac{1}{2}^{+})$&~&$\Omega_{cc}(\frac{1}{2}^{-})$&~\\
parent&2.044$\pm$0.331&13.392$\pm$0.982&1.725$\pm$0.204&15.617$\pm$0.469\\
1 daughter&1.958$\pm$0.158&16.669$\pm$0.429&1.744$\pm$0.068&18.569$\pm$0.156\\
2 daughter&2.017$\pm$0.297&17.708$\pm$0.850&1.723$\pm$0.234&19.992$\pm$0.536\\
3 daughter&1.807$\pm$0.994&19.626$\pm$0.285&1.551$\pm$0.083&21.339$\pm$0.192\\
\end{tabular}
\end{ruledtabular}
\end{table*}
where $\alpha$ and $\alpha_{0}$ are slope and intercept. The straight lines in Fig. \ref{figure6} are obtained by linear fitting of the predicted values. The fitted slopes and intercepts of the Regge trajectories are listed in Table \ref{tableII}.
We can see from these figures that all of the predicted masses in our model fit nicely to the linear trajectories in the ($J$,$M^{2}$) plane. These results can help us to assign an accurate position in the mass spectra for observed doubly charmed baryons in the future.
\section{ Conclusions}
In this work, we have systematically investigate the mass spectra, the r.m.s. radii and the radial density distributions of the doubly charmed baryons $\Xi_{cc}$, and $\Omega_{cc}$ in the frame work of relativized quark model. In addition, with the predicted mass spectra, we also construct the Regge trajectories in ($J$,$M^{2}$) plane.
The first feature of this work is that a doubly charmed baryon is regarded as a three-body system of quarks and all quarks contribute fully to the dynamics in the baryon. This is different with the light-quark-heavy-diqurak approximation where the three-body problem is reduced to two-body calculations.
Second, all parameters in our calculations such as quark masses and parameters of the interquark potential are consistent with those of our previous work\cite{GLY1,ZYL1}.
Third, it is the first time that the masses, r.m.s. radii and radial density distributions of the ground, orbital and radial excited states($1S\sim4S$, $1P\sim4P$, $1D\sim4D$, $1F\sim4F$ and $1G\sim4G$) are systematically studied(in Tables \ref{tableIII}-\ref{tableIV}). It is found that model predicted mass of $\Xi_{cc}^{+}$ $3640$ MeV is in agreement with the experimental data $3621.4$ MeV.
Finally, for the three orbital excitations $\lambda$-mode, $\rho$-mode and $\lambda$-$\rho$ mixing mode, it is shown that the mixing of these excited modes is suppressed and only $\rho$-mode dominates.

Up to now, only the ground state 1S($\frac{1}{2}^{+}$) in $\Xi_{cc}$ family has been observed and confirmed in experiments. The other states that are predicted by quark model in this work, \emph{e}.\emph{g}. the $\Xi_{cc}$ baryons with quantum numbers  1S($\frac{3}{2}^{+}$), 1P($\frac{1}{2}^{-}$), 1P($\frac{3}{2}^{-}$), and the 1S and 1P wave $\Omega_{cc}$ baryons, are still missing in experiments. These baryons are either the ground state $\Omega_{cc}$($\frac{1}{2}^{+}$) or the low-lying excitations. They all have good potentials to be observed and need to be searched for by LHCb, BarBar, Bell, CLEO, BESIII collaborations. The ground state $\Omega_{cc}$($\frac{1}{2}^{+}$) has a mass of 3750 MeV, which is below the threshold of decay channels $\Xi_{c}D$, $\Xi_{c}^{\prime}D$ and $\Xi_{cc}K$, respectively. Thus, it may be searched for in the two-body weak decays\cite{decay0,decay1} $\Omega_{cc}\rightarrow\Omega_{c}\pi$, $\Omega_{cc}\rightarrow\Xi_{c}\overline{K}$ and $\Omega_{cc}\rightarrow\Xi_{c}^{\prime}\overline{K}$. For the 1S wave $\Omega_{cc}$($\frac{3}{2}^{+}$) and $\Xi_{cc}$($\frac{3}{2}^{+}$) baryons, their masses are 3799 and 3695 MeV which are also lower than the threshold of their strong decays.
The 1P wave states of $\Xi_{cc}$ and $\Omega_{cc}$ are above the threshold of $\Xi_{cc}\pi$ and $\Xi_{cc}K$, respectively. However, it was studied that these strong decays are forbidden due to the orthogonality of spatial wave functions if the simple harmonic oscillator wave functions were adopted for the 1P and 1S states\cite{decay2}. It was suggested that those low-lying doubly charmed baryons have good potentials to be observed in radiative decays\cite{decay2,decay3}.
In summary, we hope these analyses can help to search for doubly charmed baryons in future experiments.

\begin{Large}
Acknowledgments
\end{Large}
This project is supported by National Natural Science Foundation, Grant Number 12175068 and Natural Science Foundation of HeBei Province, Grant Number A2018502124.

\clearpage
\begin{widetext}
\begin{large}
\textbf{Appendix A: Masses and r.m.s. radii of the $\Xi_{cc}$ and $\Omega_{cc}$ heavy baryons}
\end{large}
\begin{table*}[htbp]
\begin{ruledtabular}\caption{Masses(in MeV) and r.m.s. radii(in fm) of the $\Xi_{cc}$ heavy baryons}
\label{tableIII}
\begin{tabular}{c c c c c| c c c c c}
$l_{\rho}$  $l_{\lambda}$ L s j & $nL$($J^{P}$) &M &$\sqrt {\langle {r_{\rho}^{2}}\rangle }$ &$\sqrt {\langle {r_{\lambda}^{2}}\rangle }$  &$l_{\rho}$  $l_{\lambda}$ L s j & $nL$($J^{P}$) &M &$\sqrt {\langle {r_{\rho}^{2}}\rangle }$ &$\sqrt {\langle {r_{\lambda}^{2}}\rangle }$ \\ \hline
\multirow{4}{*}{0 0 0 1 1 }
~& $1S$($\frac{1}{2}^{+}$) & 3640 & 0.435 &0.462   & \multirow{4}{*}{2 0 2 1 3}  & $1D$($\frac{7}{2}^{+}$) & 4233 &0.856 &0.596 \\
~& $2S$($\frac{1}{2}^{+}$) & 4069 &0.759 &0.585                                & & $2D$($\frac{7}{2}^{+}$) & 4556 &1.200 &0.675 \\
~& $3S$($\frac{1}{2}^{+}$) & 4182 &0.524 &0.836  &~                              & $3D$($\frac{7}{2}^{+}$) & 4701 &0.932 &0.959 \\
~& $4S$($\frac{1}{2}^{+}$) & 4411 &1.147 &0.658 &~                               & $4D$($\frac{7}{2}^{+}$) & 4807 &1.316 &0.712 \\ \hline
\multirow{4}{*}{0 0 0 1 1}
~ & $1S$($\frac{3}{2}^{+}$) & 3695 &0.443 &0.496 & \multirow{4}{*}{3 0 3 0 3}    & $1F$($\frac{5}{2}^{-}$) & 4342 &0.983 &0.550 \\
~ & $2S$($\frac{3}{2}^{+}$) & 4111 &0.764 &0.623 &                               & $2F$($\frac{5}{2}^{-}$) & 4663 &1.347 &0.629 \\
~ & $3S$($\frac{3}{2}^{+}$) & 4209 &0.539 &0.851  &                             ~& $3F$($\frac{5}{2}^{-}$) & 4824 &1.097 &0.925 \\
~ & $4S$($\frac{3}{2}^{+}$) & 4445 &1.147 &0.689  &~                             & $4F$($\frac{5}{2}^{-}$) & 4891 &1.372 &0.661 \\ \hline
\multirow{4}{*}{1 0 1 0 1}
~ & $1P$($\frac{1}{2}^{-}$) & 3932 &0.642 &0.502  &\multirow{4}{*}{3 0 3 0 3}   &$1F$($\frac{7}{2}^{-}$) & 4422 &1.007 &0.625 \\
~ & $2P$($\frac{1}{2}^{-}$) & 4289 &0.946 &0.589 &                            ~ & $2F$($\frac{7}{2}^{-}$) & 4736 &1.426 &0.712 \\
~ & $3P$($\frac{1}{2}^{-}$) & 4447 &0.717 &0.886 &                           ~ & $3F$($\frac{7}{2}^{-}$) & 4876 &1.093 &0.984 \\
~ & $4P$($\frac{1}{2}^{-}$) & 4582 &1.299 &0.669  &                          ~ & $4F$($\frac{7}{2}^{-}$) & 4971 &1.298 &0.711 \\ \hline
\multirow{4}{*}{1 0 1 0 1}
~ & $1P$($\frac{3}{2}^{-}$) & 3978 &0.654 &0.536 &\multirow{4}{*}{4 0 4 1 3}  &$1G$($\frac{5}{2}^{+}$) & 4526 &1.121 &0.577 \\
~ & $2P$($\frac{3}{2}^{-}$) & 4328 &0.963 &0.622 &                           ~ & $2G$($\frac{5}{2}^{+}$) & 4843 &1.600 &0.672 \\
~ & $3P$($\frac{3}{2}^{-}$) & 4472 &0.722 &0.907 &                          ~ & $3G$($\frac{5}{2}^{+}$) & 4990 &1.233 &0.944 \\
~ & $4P$($\frac{3}{2}^{-}$) & 4615 &1.290 &0.695 &                          ~ & $4G$($\frac{5}{2}^{+}$) & 5063 &1.308 &0.652 \\ \hline
\multirow{4}{*}{2 0 2 1 1}
~ & $1D$($\frac{1}{2}^{+}$) & 4163 &0.819 &0.538&\multirow{4}{*}{4 0 4 1 3}  &$1G$($\frac{7}{2}^{+}$) & 4599 &1.142 &0.650 \\
~ & $2D$($\frac{1}{2}^{+}$) & 4490 &1.121 &0.614  & ~ & $2G$($\frac{7}{2}^{+}$) & 4905 &1.669 &0.752 \\
~ & $3D$($\frac{1}{2}^{+}$) & 4656 &0.908 &0.916 & ~ & $3G$($\frac{7}{2}^{+}$) & 5041 &1.224 &1.006 \\
~ & $4D$($\frac{1}{2}^{+}$) & 4745 &1.381 &0.679 &~ & $4G$($\frac{7}{2}^{+}$) & 5146 &1.240 &0.705 \\ \hline
\multirow{4}{*}{2 0 2 1 1}
~ & $1D$($\frac{3}{2}^{+}$) & 4203 &0.831 &0.572 & \multirow{4}{*}{4 0 4 1 4} & $1G$($\frac{7}{2}^{+}$) & 4511 &1.120 &0.565 \\
~ & $2D$($\frac{3}{2}^{+}$) & 4526 &1.148 &0.648                              &~ & $2G$($\frac{7}{2}^{+}$) & 4829 &1.599 &0.660 \\
~ & $3D$($\frac{3}{2}^{+}$) & 4679 &0.911 &0.940 &                            ~ & $3G$($\frac{7}{2}^{+}$) & 4979 &1.236 &0.936 \\
~ & $4D$($\frac{3}{2}^{+}$) & 4777 &1.359 &0.703 &                            ~ & $4G$($\frac{7}{2}^{+}$) & 5050 &1.310 &0.642 \\ \hline
\multirow{4}{*}{2 0 2 1 2}
~ & $1D$($\frac{3}{2}^{+}$) & 4151 &0.821 &0.528 &\multirow{4}{*}{4 0 4 1 4}& $1G$($\frac{9}{2}^{+}$) & 4605 &1.147 &0.659 \\
~ & $2D$($\frac{3}{2}^{+}$) & 4481 &1.124 &0.604 &                           ~ & $2G$($\frac{9}{2}^{+}$) & 4909 &1.683 &0.762 \\
~ & $3D$($\frac{3}{2}^{+}$) & 4650 &0.915 &0.910 &                           ~ & $3G$($\frac{9}{2}^{+}$) & 5045 &1.225 &1.014 \\
~ & $4D$($\frac{3}{2}^{+}$) & 4736 &1.379 &0.669&                           ~ & $4G$($\frac{9}{2}^{+}$) & 5158 &1.227 &0.713 \\ \hline
\multirow{4}{*}{2 0 2 1 2}
~ & $1D$($\frac{5}{2}^{+}$) & 4217 &0.841 &0.583 &\multirow{4}{*}{4 0 4 1 5}&$1G$($\frac{9}{2}^{+}$) & 4495 &1.122 &0.553 \\
~ & $2D$($\frac{5}{2}^{+}$) & 4540 &1.169 &0.660 &~ & $2G$($\frac{9}{2}^{+}$) & 4816 &1.603 &0.649 \\
~ & $3D$($\frac{5}{2}^{+}$) & 4689 &0.919 &0.949 &~ & $3G$($\frac{9}{2}^{+}$) & 4968 &1.238 &0.929 \\
~ & $4D$($\frac{5}{2}^{+}$) & 4790 &1.342 &0.708&~ & $4G$($\frac{9}{2}^{+}$) & 5039 &1.305 &0.630 \\ \hline
\multirow{4}{*}{2 0 2 1 3}
~ & $1D$($\frac{5}{2}^{+}$) & 4142 &0.829 &0.519 &\multirow{4}{*}{4 0 4 1 5}&$1G$($\frac{11}{2}^{+}$) & 4611 &1.154 &0.669 \\
~ & $2D$($\frac{5}{2}^{+}$) & 4474 &1.135 &0.594 &~ & $2G$($\frac{11}{2}^{+}$) & 4914 &1.700 &0.773 \\
~ & $3D$($\frac{5}{2}^{+}$) & 4647 &0.926 &0.905 &~ & $3G$($\frac{11}{2}^{+}$) & 5048 &1.228 &1.024 \\
~ & $4D$($\frac{5}{2}^{+}$) & 4729 &1.371 &0.658 &~ & $4G$($\frac{11}{2}^{+}$) & 5172 &1.214 &0.722
\end{tabular}
\end{ruledtabular}
\end{table*}
\begin{table*}[htbp]
\begin{ruledtabular}\caption{Masses(in MeV) and r.m.s. radii(in fm) of the $\Omega_{cc}$ heavy baryons}
\label{tableIV}
\begin{tabular}{c c c c c| c c c c c}
$l_{\rho}$  $l_{\lambda}$ L s j & $nL$($J^{P}$) &M &$\sqrt {\langle {r_{\rho}^{2}}\rangle }$ &$\sqrt {\langle {r_{\lambda}^{2}}\rangle }$ &$l_{\rho}$  $l_{\lambda}$ L s j & $nL$($J^{P}$) &M &$\sqrt {\langle {r_{\rho}^{2}}\rangle }$ &$\sqrt {\langle {r_{\lambda}^{2}}\rangle }$  \\ \hline
\multirow{4}{*}{0 0 0 1 1 }
~& $1S$($\frac{1}{2}^{+}$) & 3750 &0.426 &0.427   &\multirow{4}{*}{2 0 2 1 3} &  $1D$($\frac{7}{2}^{+}$) & 4346 &0.844 &0.553 \\
~& $2S$($\frac{1}{2}^{+}$) & 4182 &0.731 &0.567  & ~ & $2D$($\frac{7}{2}^{+}$) & 4671 &1.159 &0.633 \\
~& $3S$($\frac{1}{2}^{+}$) & 4291 &0.540 &0.775  &~ & $3D$($\frac{7}{2}^{+}$) & 4807 &0.932 &0.907 \\
~& $4S$($\frac{1}{2}^{+}$) & 4531 &1.120 &0.650 &~ & $4D$($\frac{7}{2}^{+}$) & 4921 &1.345 &0.682 \\ \hline
\multirow{4}{*}{0 0 0 1 1}
~ & $1S$($\frac{3}{2}^{+}$) & 3799 &0.435 &0.457 &\multirow{4}{*}{3 0 3 0 3}&$1F$($\frac{5}{2}^{-}$) & 4471 &0.973 &0.519 \\
~ & $2S$($\frac{3}{2}^{+}$) & 4219 &0.735 &0.599 &~ & $2F$($\frac{5}{2}^{-}$) & 4792 &1.302 &0.598 \\
~ & $3S$($\frac{3}{2}^{+}$) & 4315 &0.553 &0.790  &~ & $3F$($\frac{5}{2}^{-}$) & 4937 &1.094 &0.878 \\
~ & $4S$($\frac{3}{2}^{+}$) & 4561 &1.121 &0.677  &~ & $4F$($\frac{5}{2}^{-}$) & 5018 &1.414 &0.636 \\ \hline
\multirow{4}{*}{1 0 1 0 1}
~ & $1P$($\frac{1}{2}^{-}$) & 4049 &0.631 &0.468&\multirow{4}{*}{3 0 3 0 3}& $1F$($\frac{7}{2}^{-}$) & 4538 &0.995 &0.582 \\
~ & $2P$($\frac{1}{2}^{-}$) & 4407 &0.922 &0.561 &~ & $2F$($\frac{7}{2}^{-}$) & 4854 &1.377 &0.667 \\
~ & $3P$($\frac{1}{2}^{-}$) & 4557 &0.718 &0.834 &~ & $3F$($\frac{7}{2}^{-}$) & 4984 &1.098 &0.932 \\
~ & $4P$($\frac{1}{2}^{-}$) & 4706 &1.298 &0.646  &~ & $4F$($\frac{7}{2}^{-}$) & 5083 &1.343 &0.681 \\ \hline
\multirow{4}{*}{1 0 1 0 1}
~ & $1P$($\frac{3}{2}^{-}$) & 4089 &0.642 &0.497 &\multirow{4}{*}{4 0 4 1 3}&$1G$($\frac{5}{2}^{+}$) & 4658 &1.112 &0.545 \\
~ & $2P$($\frac{3}{2}^{-}$) & 4441 &0.936 &0.589 &~ & $2G$($\frac{5}{2}^{+}$) & 4976 &1.551 &0.639 \\
~ & $3P$($\frac{3}{2}^{-}$) & 4579 &0.723 &0.854 &~ & $3G$($\frac{5}{2}^{+}$) & 5104 &1.242 &0.897 \\
~ & $4P$($\frac{3}{2}^{-}$) & 4734 &1.292 &0.667 &~ & $4G$($\frac{5}{2}^{+}$) & 5188 &1.357 &0.629 \\ \hline
\multirow{4}{*}{2 0 2 1 1}
~ & $1D$($\frac{1}{2}^{+}$) & 4285 &0.807 &0.504&\multirow{4}{*}{4 0 4 1 3}&$1G$($\frac{7}{2}^{+}$) & 4718 &1.131 &0.607 \\
~ & $2D$($\frac{1}{2}^{+}$) & 4612 &1.088 &0.582  &~ & $2G$($\frac{7}{2}^{+}$) & 5028 &1.623 &0.706 \\
~ & $3D$($\frac{1}{2}^{+}$) & 4767 &0.904 &0.867 &~ & $3G$($\frac{7}{2}^{+}$) & 5148 &1.238 &0.954 \\
~ & $4D$($\frac{1}{2}^{+}$) & 4870 &1.402 &0.654 &~ & $4G$($\frac{7}{2}^{+}$) & 5255 &1.285 &0.677 \\ \hline
\multirow{4}{*}{2 0 2 1 1}
~ & $1D$($\frac{3}{2}^{+}$) & 4318 &0.818 &0.532 &\multirow{4}{*}{4 0 4 1 4}&$1G$($\frac{7}{2}^{+}$) & 4645 &1.112 &0.536 \\
~ & $2D$($\frac{3}{2}^{+}$) & 4642 &1.111 &0.610 &~ & $2G$($\frac{7}{2}^{+}$) & 4965 &1.550 &0.629 \\
~ & $3D$($\frac{3}{2}^{+}$) & 4788 &0.908 &0.889 &~ & $3G$($\frac{7}{2}^{+}$) & 5093 &1.244 &0.889 \\
~ & $4D$($\frac{3}{2}^{+}$) & 4896 &1.384 &0.674 &~ & $4G$($\frac{7}{2}^{+}$) & 5177 &1.358 &0.620 \\ \hline
\multirow{4}{*}{2 0 2 1 2}
~ & $1D$($\frac{3}{2}^{+}$) & 4276 &0.810 &0.496 &\multirow{4}{*}{4 0 4 1 4}& $1G$($\frac{9}{2}^{+}$) & 4722 &1.136 &0.615 \\
~ & $2D$($\frac{3}{2}^{+}$) & 4604 &1.091 &0.573 &~ & $2G$($\frac{9}{2}^{+}$) & 5032 &1.640 &0.715 \\
~ & $3D$($\frac{3}{2}^{+}$) & 4762 &0.910 &0.861 &~ & $3G$($\frac{9}{2}^{+}$) & 5151 &1.239 &0.963 \\
~ & $4D$($\frac{3}{2}^{+}$) & 4863 &1.400 &0.645&~ & $4G$($\frac{9}{2}^{+}$) & 5265 &1.269 &0.684 \\ \hline
\multirow{4}{*}{2 0 2 1 2}
~ & $1D$($\frac{5}{2}^{+}$) & 4331 &0.829 &0.543 &\multirow{4}{*}{4 0 4 1 5}&$1G$($\frac{9}{2}^{+}$) & 4632 &1.114 &0.526 \\
~ & $2D$($\frac{5}{2}^{+}$) & 4655 &1.130 &0.620 &~ & $2G$($\frac{9}{2}^{+}$) & 4954 &1.556 &0.620 \\
~ & $3D$($\frac{5}{2}^{+}$) & 4796 &0.917 &0.897 &~ & $3G$($\frac{9}{2}^{+}$) & 5084 &1.247 &0.882 \\
~ & $4D$($\frac{5}{2}^{+}$) & 4907 &1.369 &0.678&~ & $4G$($\frac{9}{2}^{+}$) & 5168 &1.352 &0.609 \\ \hline
\multirow{4}{*}{2 0 2 1 3}
~ & $1D$($\frac{5}{2}^{+}$) & 4269 &0.818 &0.488 &\multirow{4}{*}{4 0 4 1 5}& $1G$($\frac{11}{2}^{+}$) & 4727 &1.142 &0.622 \\
~ & $2D$($\frac{5}{2}^{+}$) & 4600 &1.101 &0.566 &~ & $2G$($\frac{11}{2}^{+}$) & 5036 &1.660 &0.725 \\
~ & $3D$($\frac{5}{2}^{+}$) & 4759 &0.922 &0.856 &~ & $3G$($\frac{11}{2}^{+}$) & 5154 &1.242 &0.971 \\
~ & $4D$($\frac{5}{2}^{+}$) & 4857 &1.392 &0.635 &~ & $4G$($\frac{11}{2}^{+}$) & 5277 &1.252 &0.692
\end{tabular}
\end{ruledtabular}
\end{table*}
\end{widetext}
\end{document}